\documentclass[fleqn,usenatbib]{mnras}

\usepackage{newtxtext,newtxmath}

\usepackage[T1]{fontenc}
\usepackage{ae,aecompl}

\usepackage{graphicx}	
\usepackage{amsmath}	
\usepackage{amssymb}	

\usepackage{psfrag}
\usepackage{placeins}
\usepackage{diagbox}
\usepackage{color}

\newenvironment{mytabular}[1][1]{
  \renewcommand*{\arraystretch}{#1}
  \tabular
}{
  \endtabular
}

\title[Large scale structure and GDM]{Using large scale structure data and a halo model to constrain Generalised Dark Matter}

\author[D. B. Thomas et al.]{
Daniel B Thomas,$^{1}$\thanks{E-mail: daniel.thomas-2@manchester.ac.uk}
Michael Kopp,$^{2}$
and Katarina Markovi\v{c}$^{3}$\\
$^{1}$Jodrell Bank Centre for Astrophysics, School of Physics \& Astronomy, The University of Manchester, Manchester M13 9PL, UK \\
$^{2}$CEICO, Institute of Physics of the Czech Academy of Sciences, Na Slovance 2, Praha 8 Czech Republic\\
$^{3}$Institute of Cosmology and Gravitation, University of Portsmouth, Dennis Sciama Building, Portsmouth P01 3FX, UK
}

\date{Accepted XXX. Received YYY; in original form ZZZ}

\pubyear{2019}

\begin{document}
\label{firstpage}
\pagerange{\pageref{firstpage}--\pageref{lastpage}}
\maketitle

\begin{abstract}
Constraints on the properties of the cosmological dark matter have previously been obtained in a model-independent fashion using the Generalised Dark Matter (GDM) framework. Here we extend that work in several directions: We consider the inclusion of WiggleZ matter power spectrum data, and show that this improves the constraints on the two perturbative GDM parameters, $c^2_s$ and $c^2_\text{vis}$, by a factor of 3, for a conservative choice of wavenumber range. A less conservative choice can yield an improvement of up to an order of magnitude compared to previous constraints. In order to examine the robustness of this result we develop a GDM halo model to explore how non-linear structure formation could proceed in this framework, since currently GDM has only been defined perturbatively and only linear theory has been used when generating constraints. We then examine how the halo model affects the constraints obtained from the matter power spectrum data. The less-conservative wavenumber range shows a significant difference between linear and non-linear modelling, with the latter favouring GDM parameters inconsistent with $\Lambda$CDM, underlining the importance of careful non-linear modelling when using this data. We also use this halo model to establish the robustness of previously obtained constraints, particularly those that involve weak gravitational lensing of the cosmic microwave background. Additionally, we show how the inclusion of neutrino mass as a free parameter affects previous constraints on the GDM parameters.
\end{abstract}





\section{Introduction}
The $\Lambda$CDM cosmological model does a good job of reproducing the current cosmological observations. In this model, the standard model of particle physics is supplemented by a cosmological constant $\Lambda$ and a dark matter particle. This dark matter particle is assumed to interact purely due to the influence of gravity and to have a negligible (initial) velocity dispersion, thus the name Cold Dark Matter (CDM). In perturbative calculations this is typically modelled as a pressure-less perfect fluid. As a result, many cosmological constraints on the dark matter density are, more correctly, constraints on the density of this pressure-less perfect fluid. More generally, CDM is evolved by solving the collision-less Boltzmann equation. This is done on large scales using cosmological perturbation theory (implemented in Boltzmann codes such as \texttt{class} and \texttt{camb}) and on smaller scales using N-body simulations and other non-linear methods.

Since we are entering the era of so-called  ``precision cosmology,'' in which many cosmological parameters have been measured with 1\% accuracy or better, it is timely to consider whether such an idealised and simple dark matter model is sufficient when analysing the data. There are many physical dark matter models that do not yield precisely CDM, for example Warm Dark Matter (WDM) \citep{DodelsonWidrow1994,Armendariz-PiconNeelakanta2014,PiattellaCasariniFabrisEtal2015} or ultra light axions \citep{HuBarkanaGruzinov2000,HlozekGrinMarshEtal2015}, which are one example of Fuzzy Dark Matter (FDM). In addition, recent work on the Effective Field Theory of Large Scale Structure (EFTofLSS) \citep{BaumannNicolisSenatoreEtal2012} shows that even an ideal CDM candidate develops a more complicated energy momentum tensor, even on linear scales, once the non-linearities that inevitably form on small scales back-react on the large scales. This causes an effective pressure and viscosity on large scales. From a non-cosmological perspective, despite a large number of direct and indirect detection experiments for dark matter, no convincing detections have been made, and many theoretically favoured regions of parameter space have been ruled out \citep{Xenon1002012,Xenon1002014,BuckleyCowenProfumo2013,OliveEtal2014,CRESST2015,LUX2015}. Thus, there are strong reasons to go beyond the simplest ways of modelling dark matter.

In \citet{KoppSkordisThomas2016}, the Generalised Dark Matter (GDM) model (first proposed in \citet{Hu1998a}) was examined in some detail, notably how it relates to particular physical models. GDM adds to the CDM energy momentum tensor a background pressure, pressure perturbation and anisotropic stress. Closure relations are then postulated to match qualitative properties of  known models, like massive neutrinos, and in order to de-correlate background and perturbative properties. GDM encompasses WDM, FDM and the EFTofLSS effects as well as other physical models, so it is sufficiently versatile for examining dark matter properties in a model independent fashion. In \citet{ThomasKoppSkordis2016}, all GDM parameters were constrained using Cosmic Microwave Background (CMB) data, supported by additional data on the cosmological expansion history (see section 4.3 in this paper and references therein for comparison to earlier works constraining partial or similar parameters to those we consider here, such as \citet{Muller2005, CalabreseMigliaccioPaganoEtal2009,XuChang2013}). The results showed no evidence for any non-CDM properties of dark matter. This was expanded on in \citet{KoppEtal2018}, where an improved freedom was given to one of the GDM parameters; this was used to demonstrate for the first time that there is no cosmological epoch where the data would favour a nonzero equation of state, and furthermore that there is no cosmological epoch where the data is consistent with zero dark matter density, thus showing the strength of the GDM approach to testing the CDM paradigm. An independent group subsequently verified \citep{KunzNesserisSawicki2016} some of the results in \citet{ThomasKoppSkordis2016}, as well as using some late time matter clustering data; we will comment further on this later in the paper. Further work constraining the GDM parameters is now ongoing by other groups, see e.g. \citet{TutusausLamineDupaysEtal2016}.

It was noted in \citet{ThomasKoppSkordis2016} that matter power spectrum data could not only improve the constraints on the GDM parameters, but also has the potential to break a degeneracy between two of them (see section \ref{sec_gdm}). The robust use of such data requires a non-linear extension of the GDM model, which is not present in the literature. It was also noted in \citet{ThomasKoppSkordis2016} that the inclusion of a non-linear extension to perturbation theory for $\Lambda$CDM makes a difference to the CMB lensing potential. This effect is of a similar magnitude, but opposite sign, to that of GDM with parameters saturating the constraints found in \citet{ThomasKoppSkordis2016}. In this paper we develop a halo model for GDM which allows us  first to test the robustness of the results in \citet{ThomasKoppSkordis2016}, and second to use matter power spectrum data from the WiggleZ survey \citep{ParkinsonEtal2012} to improve the constraints on the GDM parameters. The paper is laid out as follows: in section \ref{sec_gdm} we briefly review the GDM model and previous constraints, before constructing the GDM halo model in section \ref{sec_halo}. We then present the methodology for our constraints in section \ref{sec_method} and present our resulting constraints and robustness tests in section \ref{sec_results}. We conclude in section \ref{sec_conc}.

\section{Brief review of GDM model and previous constraints}
\label{sec_gdm}
The GDM model was first proposed as an extension to the standard CDM model in \citet{Hu1998a}. Here we give brief details of the model, following \citet{KoppSkordisThomas2016}; see both this work and \citet{Hu1998a} for further details of the model, its motivation and the different physical models that it can encompass.\\
The standard CDM energy momentum tensor is given by 
\begin{equation}
T_{\mu \nu}=\rho u_\mu u_\nu \,\text{,}
\end{equation}
i.e. the fluid is specified entirely by its density, $\rho$, and velocity $u^\mu$. This is then typically divided into a background part that is homogeneous and a perturbation. The GDM parameterisation adds pressure and anisotropic stress to this, giving
\begin{equation}
T_{\mu \nu}=(\rho+P) u_\mu u_\nu+Pg_{\mu\nu} +\Sigma_{\mu \nu} \,\text{.}
\end{equation}
The pressure and density perturbations are divided into background quantities (denoted by an overbar) and perturbed quantities as usual, and the additional scalar perturbations of $P$ and $\Sigma_{\mu \nu}$ are controlled by the equation of state $w$ (background pressure), the sound speed $c^2_s$ (pressure perturbation) and the viscosity $c^2_{\text{vis}}$ (anisotropic stress). The equation of state relates to the background quantities in the usual way: $w=\bar{\rho}/\bar{P}$, and the additional perturbations are governed by the closure equations
 \begin{eqnarray}
  \Pi &=&  c_a^2 \delta +   \left( c_s^2 - c_a^2 \right) \hat{\Delta}\\
\dot{\Sigma} &=&  - 3 {\cal{H}}\Sigma+ \frac{4}{1+w} c^2_\text{vis} \hat{\Theta} \text{.}
\end{eqnarray}
Here, $\hat{\Delta}$ and $\hat{\Theta}$ are (a particular choice of)  gauge invariant density and velocity perturbations for GDM, and $\Pi$ and $\Sigma$ are the pressure and (scalar) anisotropic stress perturbations. The adiabatic sound speed is $c^2_a=\dot{\bar{P}}/ \dot{\bar{\rho}}=w-\frac{\dot{w}}{3{\cal{H}}(1+w)}$.
Note that overdots refer to conformal time $\eta$. See \citet{KoppSkordisThomas2016} for an in depth explanation of our notation and of this choice of closure equations. Note that $w=c^2_s=c^2_\text{vis}=0$ recovers the pressureless perfect fluid, and therefore the standard $\Lambda$CDM cosmological model.

Both $c^2_s$ and $c^2_\text{vis}$ cause a decay in the gravitational potential power spectrum on scales below $k^{-1}_\text{dec}(\eta)\approx\eta\sqrt{c^2_s+\frac{8}{15}c^2_\text{vis}}$ in a GDM dominated universe \citep{KoppSkordisThomas2016}. In addition, if $c^2_s$ is sufficiently larger than $c^2_\text{vis}$, then it causes oscillations in the density perturbation below the Jeans length. Although we refer to $c^2_s$ as the sound speed, this is technically only true if $c^2_s\gg c^2_\text{vis}$ (see \citet{KoppSkordisThomas2016}). The viscosity $c^2_\text{vis}$ damps the density perturbations without causing any oscillations. As expected, the equation of state $w$ changes the expansion history of the universe for a fixed $\Omega_m$. In particular, in \citet{KoppSkordisThomas2016} it was shown that the main effect is to change the time of matter-radiation equality, and thus to change the relative heights of the peaks in the CMB. In addition, $w$ changes the distance to the last scattering surface.

The aforementioned phenomenology of the parameters is all manifest in \citet{ThomasKoppSkordis2016}. In this work we took simple forms of the GDM parameters, giving them a single value with no time and scale dependence. We then constrained these parameters using Planck CMB data (temperature, polarisation and the lensing potential) \citep{PlanckCollaborationXI2015}, BAO data \citep{BeutlerBlakeCollessEtAl2011,AndersonAubourgBaileyEtal2014} and a $H_0$ prior from the HST key project \citep{RiessMacriCasertanoEtal2011}. We found upper bounds on $c^2_s$ and $c^2_\text{vis}$ of $3.21\times10^{-6}$ and $6.06\times10^{-6}$ respectively (at the 99\% confidence level), in line with the degeneracy expected if $k^{-1}_\text{dec}$ was primarily constrained by the CMB. We also put constraints on $w$ and found degeneracies between $w$ and $H_0,\Omega_m$ due to the effects above. In all cases we found no evidence for a non-zero value of any of these parameters. The first three rows of table \ref{table_results} summarise the constraints obtained in previous work. The main conclusions of  \citet{ThomasKoppSkordis2016} were independently verified in \citet{KunzNesserisSawicki2016}. Furthermore, the assumption of a single time independent value for $w$ was relaxed in \citet{KoppEtal2018}, which showed that a non-vanishing dark matter background density is required at every epoch, thus showing the power of the GDM formalism for constraining extensions to $\Lambda$CDM.

In \citet{ThomasKoppSkordis2016} it was noted that the phenomenology of the GDM model suggests that the use of late time clustering data, such as the matter power spectrum, could significantly improve the constraints on $c^2_s$ and $c^2_\text{vis}$. In principle, if the data is precise enough to determine oscillations in the decaying region, then such data could also break the $k_\text{dec}$ degeneracy between $c^2_s$ and $c^2_\text{vis}$. However, the use of matter power spectrum data requires going beyond the scales where perturbation theory is valid. The GDM model as described in this section is only defined perturbatively and thus must be extended in order to be applicable on smaller scales. One of the main results of this paper is the development of a halo model extension to the GDM model (see section \ref{sec_halo}): As well as seeking to improve the constraints on GDM using matter power spectrum data, we also to seek to understand how \textit{safe} this process is, i.e. how robust it is to non-linear modelling. In \citet{KunzNesserisSawicki2016}, the authors used late time matter clustering data (weak lensing data) which probes the same underlying potential power spectrum as the matter power spectrum does. However, they did not consider a non-linear modelling of the GDM in that work; instead they used \textit{halofit} \citep{SmithEtal2003} as a sanity check, whilst noting themselves that \textit{halofit} has limitations  when applied outside of a $\Lambda$CDM context. Our goal is thus not just to improve the constraints on the GDM parameters using matter power spectrum data, but also to be able to quantify how much we can trust any such results.

\begin{figure}
  \centering
  \includegraphics[width=\columnwidth]{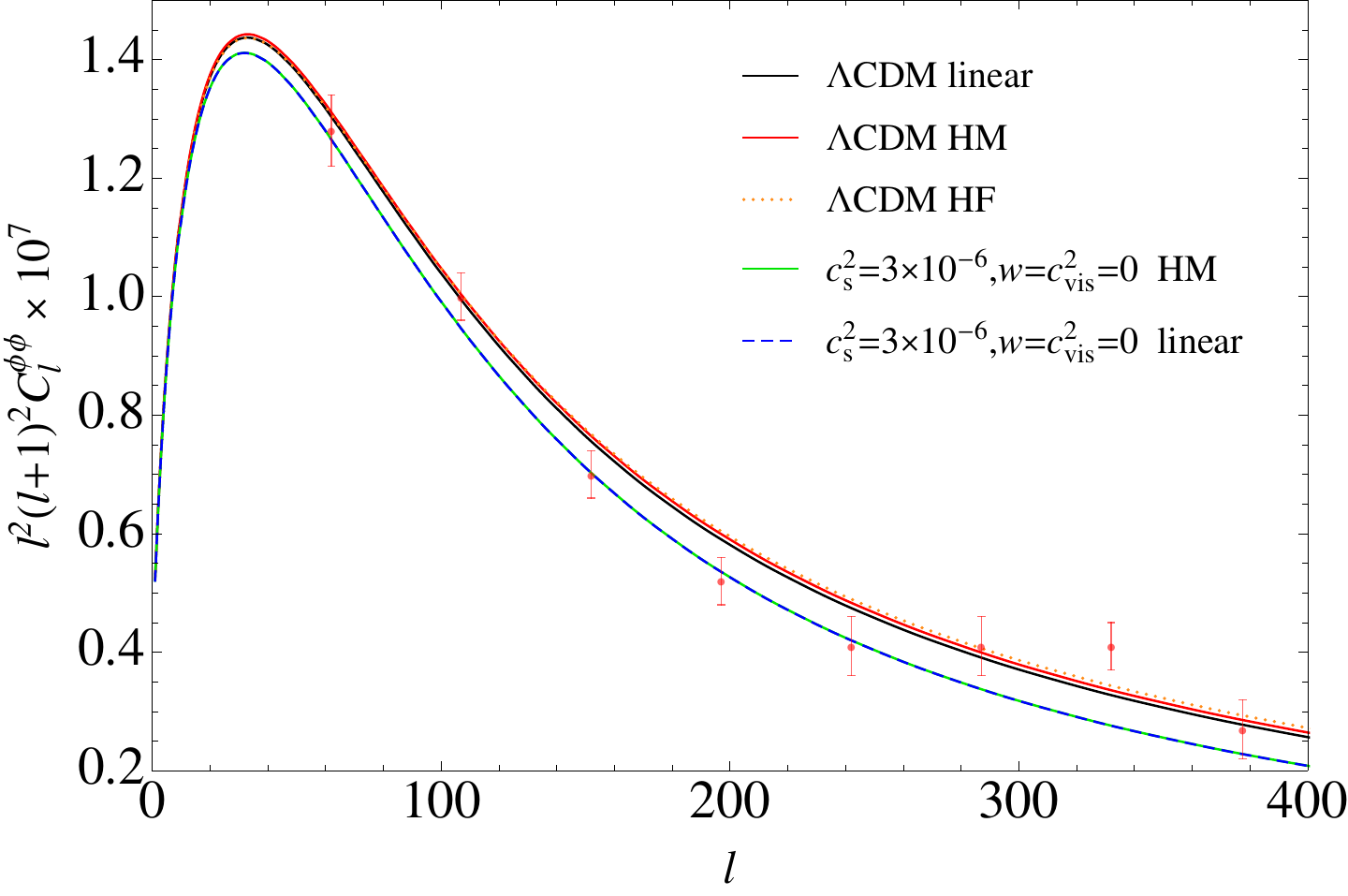}\\
  \includegraphics[width=\columnwidth]{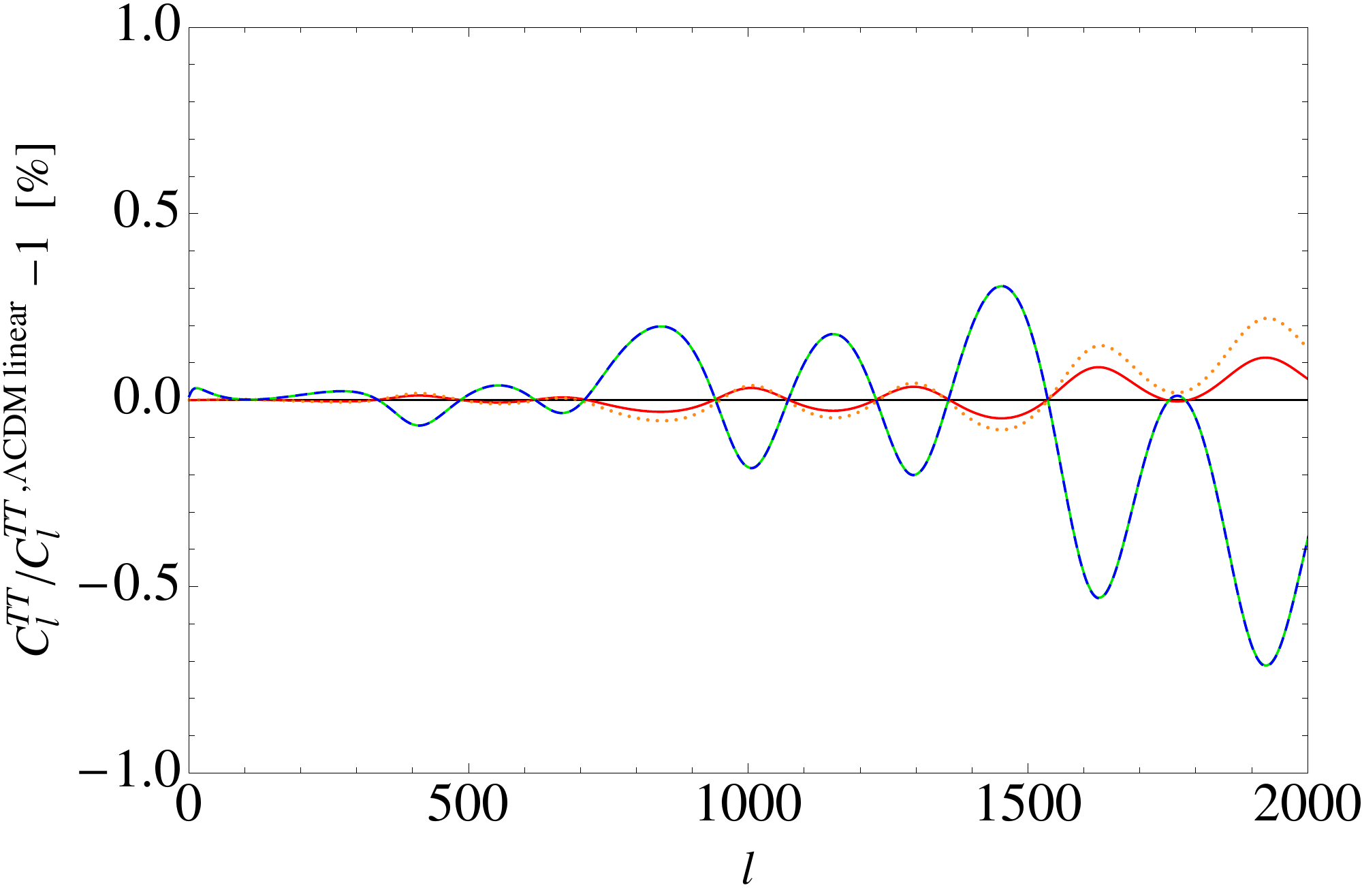}
\caption{Effect of the GDM parameters and non-linear prescriptions on the $\Lambda$CDM lensing. The upper panel shows the lensing potential and the lower panel shows the fractional change to the lensed temperature spectrum ($C^{TT}_\ell/C^{TT, \Lambda \mathrm{CDM}\text{ linear}}_\ell-1$). In both panels, the black curve shows the linear $\Lambda$CDM spectrum, orange (dotted) is $\Lambda$CDM with \textit{halofit}, red is $\Lambda$CDM with our halo model, blue (dashed) is linear GDM ($c^2_s=0.000003$) and green is GDM ($c^2_s=0.000003$) with our halo model. The data points with errors in the upper hand panel correspond to the Planck data. There are several important points to note here. Firstly, the halo model has a significantly smaller effect on GDM than on $\Lambda$CDM. Secondly, the effect of GDM and non-linear prescriptions for $\Lambda$CDM spectrum are opposite (and of a similar order of magnitude for the lensed temperature spectrum). We also note that there are some small differences between our halo model and \textit{halofit} (as is expected for the halo model \citep{SmithEtal2003}).
}
\label{fig_lensingphi}
\end{figure}

A further goal is to investigate the robustness of the constraints obtained in \citet{ThomasKoppSkordis2016}. More precisely, it was noted that even in $\Lambda$CDM, using \textit{halofit} makes a small difference to the lensing potential and thus to the lensed temperature and polarisation $C_\ell$s. We thus wish to determine whether inclusion of a non-linear prescription for GDM would strengthen or weaken the constraints previously obtained. Nonlinearities typically act to increase the matter power spectrum, which is the opposite effect to that caused by increasing GDM parameters, see figure \ref{fig_lensingphi}. Hence we expect that the constraints on $c^2_s$ and $c^2_\text{vis}$ could be weakened once nonlinearities are included.

Since we are focussing on the complexities introduced by the non-linearities and additional datasets, we will work with single, constant values of the GDM parameters as in \citet{ThomasKoppSkordis2016}. In particular, the assumption of time independence means that $c^2_a=w$.

\section{GDM halo model}
\label{sec_halo}
As stated in section \ref{sec_gdm}, the GDM model is only defined for linear perturbations and the homogeneous background, and thus cannot be constrained by all of the currently available cosmological data. One framework that has been used to make predictions on non-linear scales is the halo model \citep{CooraySheth2002}. The halo model is a semi-analytic method for computing the matter power spectrum on non-linear scales that works from the premise that the matter is organised into halos $\rho_m(\mathbf{x}) = \sum_i \rho_{\rm halo}(\mathbf{x} - \mathbf{x}_i)$ and that averaging over all of the halos gives the mean matter density in the universe $\langle \rho_m(\mathbf{x}) \rangle = \int dM M \frac{dn}{dM} = \bar{\rho}_m$, where $\frac{dn}{dM} $ is the halo mass function. The two point correlation of the matter field will thus depend on the halo density profile and also the halo mass function. For the former we will use the empirical Navarro-Frenk-White (NFW) profile and for the latter excursion set theory to predict the so-called multiplicity function $f(\sigma)$ and relate it to the halo mass function through $\frac{dn}{d \ln \sigma^{-1}}=\frac{\bar{\rho}}{M}f(\sigma)$.

\begin{figure}
  \centering
 \includegraphics[width=\columnwidth]{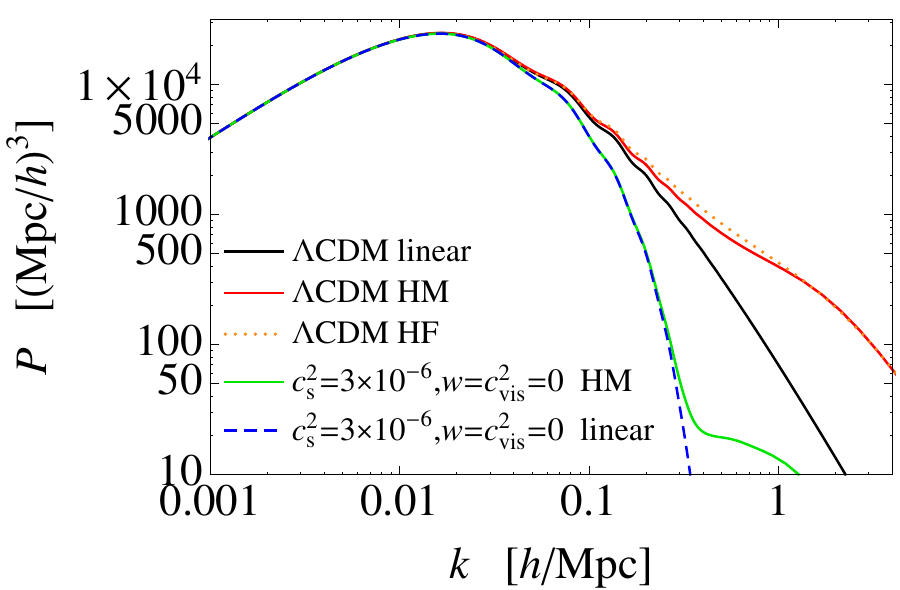}
\caption{Matter power spectra for different cosmologies and non-linear prescriptions. Black is linear $\Lambda$CDM, orange (dotted) is $\Lambda$CDM with \textit{halofit}, red is $\Lambda$CDM with our halo model, blue (dashed) is linear GDM ($\Lambda$CDM with $c^2_s=0.000003$) and green is GDM  ($\Lambda$CDM with $c^2_s=0.000003$) with our halo model. For GDM, the non-linear effects do not become significant until smaller scales than for $\Lambda$CDM, but the difference between the GDM linear and non-linear spectra increases much more sharply. The difference between the cosmologies is dominated by the linear theory decay in GDM. Also note the small difference between \textit{halofit} and the halo model for $\Lambda$CDM.}
\label{fig_spectra}
\end{figure}

For more details we refer to appendix \ref{sec_LCDMhalo}, where we present the $\Lambda$CDM halo model \citep{Seljak2000}. This also serves to introduce our notation and perspective, as these can vary between presentations of the halo model. We also present our mass function in this appendix. There is another commonly used non-linear correction for $\Lambda$CDM: \textit{halofit} \citep{SmithEtal2003}, which is an extension of the halo model that is calibrated against N-body simulations. As was already noticed in \citet{KunzNesserisSawicki2016}, it does not make sense to use \textit{halofit} for GDM, as GDM is not part of the cosmologies that have been used to calibrate it. Furthermore, the numerical implementation in Boltzmann codes (e.g. \texttt{class}) simply crashes for values of GDM parameters where the linear power spectrum falls off too quickly. See figures \ref{fig_lensingphi} and \ref{fig_spectra} for the differences between \textit{halofit} and the halo model as implemented in this paper for a $\Lambda$CDM cosmology; it is known that \textit{halofit} and the halo model differ for $\Lambda$CDM \citep{SmithEtal2003}. This difference is largest, up to 15\%, in the interval $0.1 <k [h/\mathrm{Mpc}] < 1$, visible in figure \ref{fig_spectra} comparing the red and dotted orange lines.  However, for $ k< 0.1 h/\mathrm{Mpc}$, relevant for our applications the agreement between our halo model and \textit{halofit} is better than 2\% and sufficient.

Examining the standard $\Lambda$CDM halo model, we can see that there are several obstacles to simply applying the framework ``as is'' to GDM. Firstly, due to the large drop in power on small scales when $c^2_s$ and $c^2_v$ are non-zero, the equation defining $M_\star$ does not necessarily have a solution. In addition, a lot of the formulas that are used are calibrated against $\Lambda$CDM N-body simulations. We thus implement the halo model for GDM using a similar approach to the \texttt{Warm and Fuzzy} code \citep{Marsh2016}, which is designed for warm dark matter and axions (note that both of these can fit into the GDM parameterisation \citep{KoppSkordisThomas2016}). In this approach, four modifications are made relative to a $\Lambda$CDM cosmology: A) a modified linear spectrum, B) a modified halo concentration, C) a mass dependent barrier related to critical density for spherical collapse, and D)  a modified mass function to account for the non-Markovian corrections to standard Press-Schechter expression. We implement these corrections differently to the \texttt{Warm and Fuzzy} code; we detail our implementation here. Readers not interested in the construction and definition of our halo model can skip to section \ref{sec_method}.

\subsection{Modified linear spectrum} 
The simplest and most obvious modification is that the appropriate $\Lambda$-GDM linear theory power spectrum is used as an input for the halo model, rather than a $\Lambda$CDM spectrum. While this is done via fitting functions for the transfer functions in \texttt{Warm and Fuzzy}, we instead directly use the output from the full Boltzmann calculation from the modified \texttt{class} code.

\subsection{Modified concentration}
Following \citet{Marsh2016}, which itself follows \citet{SchneiderSmithMaccioEtal2012} for WDM, we calculate the $\Lambda$CDM value of the concentration. We then apply a correction according to
\begin{equation}
c_\text{GDM}=c_\text{$\Lambda$CDM}\left(1+\gamma_2\frac{M_{1/2}}{M} \right)^{-\gamma_2}\text{,}
\end{equation}
where $\gamma_1=15$, $\gamma_2=0.3$ and $M_{1/2}$ is the half-mode mass defined by
\begin{eqnarray} \label{halfmodemass}
M_{1/2}&=&\frac{4\pi \bar{\rho}}{3}\left( \frac{\pi}{k_\text{1/2}}\right)^3 \\
\sqrt{\frac{P_\text{GDM}(k_\text{1/2})}{P_\text{$\Lambda$CDM}(k_\text{1/2})}}&=&0.5 \text{.}
\end{eqnarray}
Note that this functional form and the specific values were found to give good fits to FDM and WDM simulations. As GDM contains these two models as limiting cases we adopt this ``as is'' to GDM. The results are not sensitive to this guess. Furthermore, in the absence of non-perturbatively defined GDM model and cosmological GDM simulations, this is the best and most conservative choice we can make, as it is known to work in two limiting cases of the GDM model.

\subsection{Mass dependent spherical collapse density $\delta_\text{crit}$ and how it relates to the mass function}
\label{sec_gdmdeltacrit}
The central object in the excursion set theory is the so-called multiplicity function $f(\sigma)$, determined by the first upcrossing rate of the smoothed random field $\delta(x, R)$ through a barrier as a function of the smoothing scale $R$. These crossings are then identified with proto-halos of size $R$, and thus fixed mass $M$, so that $f(\sigma)$ determines the mass function $dn/dM$. In its most rudimentary form that mass function is mostly sensitive to the ratio
\begin{equation}
\label{eq_timedepdcrit}
\frac{\delta_c(z, z_{\rm ini}, R)}{\sigma_R(z_{\rm ini})} \,,
\end{equation}
where $\sigma_R(z_{\rm ini})$ is the standard deviation, Eq.\,\eqref{sigma}, of the (non-relativistic and participating in structure formation) matter perturbations smoothed on a scale $R$, and $\delta_c(z, z_{\rm ini}, R)$ is the spherical collapse barrier. The redshift $z_{\rm ini}$ is a time where linear perturbation theory applies to all scales $R$ of interest and thus the statistics of the density field is gaussian, but well after radiation domination (i.e. such that a spherical collapse threshold can be obtained neglecting the radiation component). 

It is important to note that the $z$-dependence in $\delta_c$  is determined by non-perturbative fluid dynamics, it is not a density field, or some linear extrapolation of it. Rather, $\delta_c$ assigns a collapse redshift $z$ to each region (assumed for simplicity to be spherically symmetric tophat profiles) of the gaussian density field filtered at scale $R$ at time $z_{\rm ini}$; this collapse redshift $z$ is then identified with the formation time of the halo of mass $M(R)$. This assignment involves some linear dynamics (at times not much later than $z_{\rm ini}$) when the field is still linear, but more importantly it includes the fully non-perturbative collapse that defines the collapse redshift at the time $z$ when formally the density contrast $\delta(z)  \rightarrow \infty$. This formal divergence is then associated with the time of halo formation, and thus approximated as instantaneous event.

\subsubsection{Standard $\Lambda$CDM linear extrapolation}
Typically, the mass function is not written in terms of equation \eqref{eq_timedepdcrit} but instead in terms of
\begin{equation}
\frac{\delta_\text{crit}}{\sigma_R(z)} \,,
\end{equation}
where
\begin{equation} \label{eq_deltacritLCDM}
\delta _{\rm crit} := \delta_c(z, z_{\rm ini}, R) \frac{D(z)}{D(z_{\rm ini})} \text{.}
\end{equation}
 This ``linear extrapolation'' from $z_{\rm ini}$ to $z$ is possible, because in a purely CDM and $\Lambda$ dominated universe the growth $D(z)$ is scale independent. Thus,
 \begin{eqnarray}
 \label{eq_strongtimedep}
&&\frac{\delta^{\Lambda \rm CDM} _c(z, z_{\rm ini}, R)}{\sigma_R(z_{\rm ini})}
 = \frac{\delta^{\Lambda \rm CDM} _c(z, z_{\rm ini}, R) D(z)/D(z_{\rm ini})}{\sigma_R(z_{\rm ini}) D(z)/D(z_{\rm ini})} \nonumber\\
 &&= \frac{\delta^{\Lambda \rm CDM} _c(z, z_{\rm ini}, R) D(z)/D(z_{\rm ini})}{\sigma_R(z)} = \frac{\delta^{\Lambda \rm CDM} _{\rm crit}}{\sigma_R(z)}\,.
\end{eqnarray}
Writing things in this way in $\Lambda$CDM is convenient because it happens (see \citet{Weinberg2008CosmologyBook}, Chapter 8.2) that the linearly extrapolated $\delta_c$, $\delta_{\rm crit}$, is a constant  $ \delta_{\rm crit}^{\rm EdS}=\frac{3}{20} (12 \pi)^{2/3} \simeq 1.686$ in an EdS universe, and is only very mildly dependent on $z$ (but still independent of $R$) in a $\Lambda$CDM dominated universe. More precisely
\begin{align}
\delta_{\rm crit}^{\Lambda \rm CDM} &\simeq  \frac{3}{20} (12 \pi)^{2/3}(1+0.012299 \log_{10}(\Omega_m(z))) \label{eq_defDeltacritLCDM}\,\text{.}
\end{align}
This is the reason why it is common in the literature to use this linearly extrapolated $\delta_c$ (denoted here by $\delta_{\rm crit}$) and the variance $\sigma_R(z)$ of a fictitious linearly extrapolated density field, even though there is no physical interpretation for such an extrapolation. In order to more clearly separate linear from non-linear physics, and to make the least amount of guessing to arrive at our GDM mass function, we avoid here using the linearly extrapolated $\delta_c$, $\delta_{\rm crit}$, and let the excursion set theory unfold at $z_{\rm ini}$. { In other words, we are explicitly separating the purely linear effects of the GDM parameters on structure formation due to the initial density field in which halos begin to form (i.e. the accumulated changes to $P(k, z_{\rm ini})$ from times $z>z_{\rm ini}$), from the expected changes during the non-linear stages of collapse ($z<z_{\rm ini}$). These latter effects are incorporated in the changes to the spherical collapse density $\delta^{\Lambda \rm GDM}_c(z, z_{\rm ini}, R)$ in the next section (\ref{sec_gdm_deltacrit}).}

In order to proceed this way we rewrite the multiplicity function $f(\sigma)$, defined in Eq.\,\eqref{massfunc},  appearing in the mass function (see appendix \ref{sec_LCDMhalo}) using \eqref{eq_strongtimedep}. The relevant term appearing in the mass function is $\bar{B}/\sigma(z)$, see equation \eqref{asphericalBarierAchizcollapse}. Inspired by equations \eqref{eq_timedepdcrit} and \eqref{eq_strongtimedep} we multiply numerator and denominator by $\sigma(z_\text{ini})$, and define
\begin{eqnarray}
\bar{B}'=\frac{\sigma(z_\text{ini})}{\sigma(z)}\bar{B}=\delta^{\Lambda \rm CDM}_c(z, z_{\rm ini}, R)+\beta\frac{\sigma(z_\text{ini})}{\sigma(z)}\nonumber\\
=\delta^{\Lambda \rm CDM}_c(z, z_{\rm ini}, R)+\beta' \sigma^2(z_\text{ini})\text{,}
\end{eqnarray}
where $\beta'=\beta \sigma(z)/\sigma(z_\text{ini})$, the barrier is now written in terms of the ``strongly time-dependent'' spherical collapse barrier $\delta_c(z, z_{\rm ini}, R)$ and $z_\text{ini}$, and in $f(\sigma)$, $\bar{B}/\sigma(z)$ is replaced with $\bar{B}'/\sigma(z_\text{ini})$, making it manifest that the excursion set theory is applied to the random field at $z_{\rm ini}$. The Markovian part $f_0$ (see later) of the multiplicity function $f$ of the mass function is thus given by \eqref{MarkovianAchizcollapse}, replacing $\sigma(z)$ with $\sigma(z_\text{ini})$ and $\bar{B}$ with $\bar{B}'$,
\begin{align} \label{MarkovianAchizini}
f_{0}(\sigma(z_{\rm ini}),z)&=\frac{\bar B'- \sigma^2(z_{\rm ini}) d \bar B'/ d\sigma^2(z_{\rm ini})}{\sigma(z_{\rm ini})}\sqrt{\frac{2 a_b}{\pi}}e^{-\frac{a_b}{2} \left(\frac{\bar{B}'}{\sigma(z_{\rm ini})}\right)^2} \,\text{.}
\end{align}
At this point, this mass function is generic and is not derived in the context of any particular extension to $\Lambda$CDM cosmology.

To consider how this mass function relates to the standard CDM case, note that if the $d\bar{B}'/d\sigma^2$ term is ignored then this mass function reduces exactly to equation \eqref{MarkovianAchizcollapse}, just without the assumption of scale independent growth. This re-writing thus makes it clearer how scale dependent growth should manifest in the formalism. For the case of scale independent growth, the new derivative term $d\bar{B}'/d\sigma^2$ in \eqref{MarkovianAchizini} reduces to the previous form
\begin{equation}
\label{eqn_barrier}
\bar{B}'-\sigma^2_R(z_\text{ini})\frac{d\bar{B}'}{d\sigma^2_R(z_\text{ini})}=\delta^{\Lambda \rm CDM}_c = \delta^{\Lambda \rm CDM}_\text{crit}\frac{\sigma_R(z_\text{ini})}{\sigma_R(z)}\text{,}
\end{equation}
thus the mass function reduces exactly to the standard form for the $\Lambda$CDM case.

\subsubsection{GDM approach}
\label{sec_gdm_deltacrit}

We will now postulate the spherical collapse barrier $\delta^{\Lambda \rm GDM}_c$ for GDM by reversing the direction of definition in equation \eqref{eq_strongtimedep}. While in $\Lambda$CDM, $\delta_{\rm crit}^{\Lambda \rm CDM}$ is defined via \eqref{eq_strongtimedep}, leading to \eqref{eq_deltacritLCDM}, we now assume for $\Lambda$GDM the validity of equation \eqref{eq_strongtimedep} 
\begin{equation}
 \label{eq_strongtimedep_DefGDM}
\delta^{\Lambda \rm GDM}_c(z, z_{\rm ini}, R)
 :=  \delta_{\rm crit}^{\Lambda \rm CDM}  \frac{\sigma_R(z_{\rm ini})}{\sigma_R(z)}\,
\end{equation}
but use it to define $\delta^{\Lambda \rm GDM}_c(z, z_{\rm ini}, R)$ while fixing $\delta_{\rm crit}^{\Lambda \rm CDM}$ on the right hand side to be given by \eqref{eq_defDeltacritLCDM}.   
In the following we will drop the superscript $\Lambda$GDM and simply write $\delta_c(z, z_{\rm ini}, R)$ for the GDM spherical collapse barrier. The idea behind equation \eqref{eq_strongtimedep_DefGDM} is that this definition implements the intuitive idea that if in GDM power is removed in a scale dependent way at $z<z_{\rm ini}$ then the collapse should be inhibited compared to CDM, and thus the spherical collapse barrier should be increased. It is at this point unclear how to judge whether the threshold defined in this way is correct given the absence of spherical collapse simulations within a (still to be) non-linearly defined GDM model. However, in addition to increasing the threshold whenever power is removed in GDM compared the CDM, this definition smoothly and naturally reduces to the $\Lambda$CDM prescription in the CDM limit of GDM, and can thus be considered conservative for small GDM parameters.
The mean barrier for collapse at $z$ in $\Lambda$CDM can be approximated by (and we postulate this to hold in GDM too)
\begin{equation} \label{GDMmeanBarrier}
\bar B'(z , z_{\rm ini}, R) =\delta_c(z , z_{\rm ini}, R) + \tilde \beta(z) \sigma^2_R(z_{\rm ini})
\end{equation}
where $\beta(z)$ is a purely time dependent function. The multiplicity function \eqref{MarkovianAchizini} takes into account that the mean barrier $\bar B$ for a randomly selected point deviates from the spherical collapse barrier $\delta_{\rm c}$ because collapse is not spherically symmetric, and also that the barrier is diffusive, rather than 100\% absorbing. The former is parameterized by $\beta$, the latter by $a_b$. We saw above that a consistent choice in the context of scale dependent growth is $\beta'(z)=\beta\sigma(z)/\sigma(z_\text{ini})$. A conservative choice for the GDM mass function thus is
\begin{align}
\tilde \beta &=  \beta \frac{\tilde D(z)}{\tilde D(z_{\rm ini})}  \\
\tilde D(z) &\equiv \sigma_{R_{\rm max}}(z) \,,
\end{align}
where $R_{\rm max}$ is the largest smoothing scale used in the halo model code, and for all reasonably small values of GDM parameters, $\tilde{D}$ reduces to the growth function $D$. We remove the scale dependence from $\tilde {\beta}$ for two reasons. Firstly, it is not clear what it would mean to include scale dependence here; it is not done in any version of the excursion set that we know of. In addition, we wish our halo model to act like a standard $\Lambda$CDM halo model in the limit that the GDM parameters are zero. As we are working directly from a Boltzmann code and not performing the standard linear extrapolation to redshift zero, the radiation component of the universe will cause a scale dependent growth \textit{even in} $\Lambda$CDM. Thus we explicitly remove the scale dependence from this term and use a growth factor that is defined to be scale-independent even in GDM.

In order to calculate the mass function for GDM (or any other cosmology with scale-dependent growth), we need to evaluate the $d\bar{B}'/d\sigma^2$ term. The result is
\begin{equation}
\frac{d \delta_c}{d\sigma^2(z_\text{ini})} = \delta_c \frac{1}{2}\left(1- \frac{ \sigma^2_{R}(z_{\rm ini})}{ \sigma^2_{R}(z)} \frac{ d\sigma^2_{R}(z)/dR}{ d\sigma^2_{R}(z_{\rm ini})/dR}\right) \text{.}
\end{equation}
Thus, the moving barrier term is given by
\begin{equation}
\bar{B}'-\sigma^2(z_\text{ini})\frac{d\bar{B}'}{d\sigma^2(z_\text{ini})}= \delta_c \frac{1}{2}\left(1+\frac{ \sigma^2_{R}(z_{\rm ini})}{ \sigma^2_{R}(z)} \frac{ d\sigma^2_{R}(z)/dR}{ d\sigma^2_{R}(z_{\rm ini})/dR}\right) \text{,}
\end{equation}
which reduces to $\delta_c$  in the case where the initial and final $\sigma_R$ have the same shape. That is often taken to be exactly correct in a $\Lambda$CDM universe. More correctly, it depends on the value of $z_{\rm ini}$: for example, if $z_{\rm ini} =50$ since there is still enough radiation ``contamination'' left to modify the shape of the matter transfer function, such that the ratio $\frac{\sigma_R(z_{\rm ini})}{\sigma_R(z)}$ is not $R$ independent. The modification of the mass function is however less then $1\%$ for $\Lambda$CDM and can be neglected, see figure \ref{fig_sigmaratio}. However for GDM, where in general growth is scale dependent even during matter domination, we do not expect the second term in the bracket to be close to 1.

We will assume that GDM does not change the values of $\beta$ and $a_b$, which amounts to the assumption that collapse inhibition from asphericity and the scatter of the barrier due to environmental and stochastic processes are unchanged. In principle, these quantities could be measured in FDM and WDM simulations. We are not aware of any such measurements and so using the $\Lambda$CDM values seems to be the most sensible approach.

\subsection{Non-Markovian corrections and asphericity of collapse}
So far we only looked at the mass and time dependent spherical part of the collapse barrier. Now we turn to the mass and time dependent contributions to the barrier due to the asphericity of the collapse as well as the non-Markovian corrections to the mass function. For GDM, the Markovian part of the mass function, \eqref{MarkovianAchizcollapse} is replaced by \eqref{MarkovianAchizini}.

The non-Markovian corrections will be implemented in a similar fashion as done in \citet{KoppApplebyAchitouvEtal2013}. There a mass dependent spherical collapse barrier was obtained due to a modification of gravity that left the background cosmology unchanged. It was shown that the non-Markovian corrections could be included through a simple relation $f(\sigma) = f_0(\sigma) f^{\rm \Lambda CDM}(\sigma)/f_0^{\rm \Lambda CDM}(\sigma)$, where $f^{\rm \Lambda CDM}(\sigma)$ includes the known and calculable non-Markovian corrections in $\Lambda$CDM.
In our case we cannot use as reference $\Lambda$CDM because the background might be different in GDM due to $w$. 
For that reason we will define another non-Markovian reference mass function $f^{\rm ref}$, using a mass-independent spherical collapse barrier closer to $\delta_{\rm c}^{\rm GDM}$. 
We choose that reference mass-independent spherical collapse barrier to be $\delta_{\rm c, max}=\delta_{\rm c}^{\rm GDM}(M_{\rm max})$, i.e. the GDM spherical collapse barrier evaluated at the largest mass scale $M_{\rm max}$ used in the halo model code. The reference mass function thus corresponds to a fictitious GDM model in which the spherical collapse barrier is mass independent. The reason for choosing $M_{\rm max}$ is that this barrier will be similar to that of a GDM model with $c_s^2=c_{\rm vis}^2=0$ (since in the limit $k\rightarrow 0$ the effects of $c_s^2$ and $c_{\rm vis}^2$ disappear). This way we don't need to run \textit{class} twice for each model. The reference multiplicity function is  given by
\begin{equation}
f^{\rm ref}(\sigma)=f^{\rm ref}_0(\sigma)+f_{1,\tilde \beta=0}^{m-m}(\sigma)+
f_{1,\tilde \beta^{(1)}}^{m-m}(\sigma)+f_{1,\tilde \beta^{(2)}}^{m-m}(\sigma)\,,\label{ftot}
\end{equation}
where 
\begin{align*}
f_0^{\rm ref}(\sigma) &=\frac{\delta_{\rm c, max}}{\sigma}\sqrt{\frac{a_b}{2\pi}}e^{-\frac{a_b}{2} \left(\frac{\delta_{\rm c, max} + \tilde \beta \sigma^2}{\sigma}\right)^2} \\
f_{1,\beta=0}^{m-m}(\sigma)&=-\kappa a_b\dfrac{\delta_{\rm c, max}}{\sigma}\sqrt{\frac{2a_b}{\pi}}\left[\exp\left[-\frac{a_b \delta_{\rm c, max}^2}{2\sigma^2}\right]-\frac{1}{2} \Gamma\left(0,\frac{a\delta_{\rm c, max}^2}{2\sigma^2}\right)\right]\\
f_{1,\tilde \beta^{(1)}}^{m-m}(\sigma)&=- a_b\,\delta_{\rm c, max}\,\tilde \beta\left[\kappa a_b\,\text{Erfc}\left( \delta_{\rm c, max}\sqrt{\frac{a_b}{2\sigma^2}}\right)+ f_{1,\tilde \beta=0}^{m-m}(\sigma)\right]\\
f_{1,\tilde \beta^{(2)}}^{m-m}(\sigma)&=-a_b\,\tilde \beta\left[\frac{\tilde \beta}{2} \sigma^2 f_{1,\tilde \beta=0}^{m-m}(\sigma)+\delta_{\rm c, max} \,f_{1,\tilde \beta^{(1)}}^{m-m}(\sigma)\right] \\
\delta_{\rm c, max} & \equiv  \delta_c(z,z_{\rm ini},R_{\rm max}) \\
\kappa &= 0.465 \\
a_b &= 0.7143 \\
\beta & = 0.12\,.
\end{align*} 

We will also include a further correction to the mass function that has been observed to fit mass functions measured in warm dark matter simulations \citep{SchneiderSmithMaccioEtal2012,Marsh2016}. The origin of that correction is likely to also be non-Markovian in nature, and it arises due to the absence of power below the scale $k^{-1}_{\rm dec}$.
If the density field does not perform a random walk as function of $R$ it can happen that the mass function suffers a cutoff, see \citet{ParanjapeLamSheth2012}. 
If the power is sharply dropping for scales $R < k^{-1}_{\rm dec}$, then the density field at a fixed point no longer performs a random walk for varying $R$ for $R < k^{-1}_{\rm dec}$. 
Thus we expect the mass function to be more non-Markovian for those small scales, implying a cutoff determined by mass scale related to $k^{-1}_{\rm dec}$. 
We follow the fit of \citet{SchneiderSmithMaccioEtal2012}, which works well for WDM.

The final expression for the multiplicity function entering \eqref{massfunc} then is
\begin{align}
\label{eqn_fullgdmmultiplicity}
f^{\rm GDM} =\left(1+ \frac{M_{1/2}}{M} \right)^{-0.6}\,  \frac{f^{\rm ref}}{f_0^{\rm ref}}\,f_0^{\rm GDM}
\end{align}
where $M_{1/2}$, the half mode mass \eqref{halfmodemass}, is used instead of $M(k^{-1}_{\rm dec})$, see \citet{SchneiderSmithMaccioEtal2012}.
To summarize: the first two factors take into account non-Markovian effects. The first one that the random walk below $k^{-1}_{\rm dec}$ is highly non-Markovian, the second one the standard non-Markovian corrections for a diffusive barrier of the form const$_1$+const$_2 \sigma^2$. The last factor is the Markovian mass function for a moving diffusive barrier \eqref{GDMmeanBarrier}. As part of the halo model, this provides a non-linear prescription for computing the GDM matter power spectrum and comprises one of the main results of this paper. When the GDM parameters are zero, the halo model reduces to a $\Lambda$CDM halo model as expected.

We make one final modification to the $\Lambda$CDM prescription in appendix \ref{sec_LCDMhalo}: In order to apply the compensation for the 1-halo term (see appendix \ref{sec_LCDMhalo}) in GDM, we need to define a scale independent growth for GDM. In the code, we do so by replacing $\sigma_8(z)$ with $\sigma_8(z=0)\frac{\tilde{D}(z)}{\tilde{D}(z=0)}$, where $\tilde{D}(z)$ is defined to be the growth on the scale $R_{\rm max}$, which corresponds to the largest mass value computed by the code inside the halo model routine. This is chosen to be consistent with the definition of $\tilde{\beta}$.

{ In figure \ref{fig_hmf_z0}, we show the full halo mass function described here, for $\Lambda$CDM and also for several choices of constant GDM parameters.}

\subsection{$\Lambda$CDM reference model for GDM halo model}
In order to implement the halo model as described above, we need a reference $\Lambda$CDM cosmology when \texttt{class} is run for GDM (this is used to calculate the half mode mass; see equation \ref{halfmodemass}). For this, we use the Eisenstein-Hu fitting formula, for a cosmology with the same $\Omega_m$, $\Omega_b$, $\Omega_\Lambda$, $n_s$ and $H_0$ as the GDM cosmology. The two key references for the fitting formulas are \citet{EisensteinHu1998} and \citet{EisensteinHu1997}. The first of these includes the effect of baryon oscillation but not neutrinos, whereas the second takes an average over the oscillations but includes the damping effects of massive neutrinos. We implement the former of these, as used for \texttt{HMCODE} \citep{MeadPeacockHeymansEtal2015} to calculate the $\Lambda$CDM spectrum at the $k$- and $z$- values required in \texttt{class}, which are then stored in an array. This spectrum is normalised to have the same value as the \texttt{class} GDM spectrum at $k_\text{ref}$, as this scale should be above the scales that are affected by GDM.

The $\Lambda$CDM growth function from \citet{LahavLiljePrimackEtal1991} is used as part of the Eisenstein-Hu formulas,
\footnotesize
\begin{eqnarray}
D(z)&=&\frac{1+z_\text{eq}}{1+z}\frac{5}{2}\Omega_m(z)\left(\Omega_m(z)^{4/7}-\Omega_\Lambda(z)\right.\nonumber\\
&&\left.+\left(1+\frac{\Omega_m(z)}{2} \right)\left(1+\frac{\Omega_\Lambda}{70} \right) \right)^{-1}\\
\Omega_m(z)&=&\frac{\Omega_m(1+z)^3}{\Omega_m(1+z)^3+\Omega_\Lambda}\\
\Omega_\Lambda(z)&=&\frac{\Omega_\Lambda}{\Omega_m(1+z)^3+\Omega_\Lambda} \text{.}
\end{eqnarray}
\normalsize
Here, $z_\text{eq}$ is a parameter from the Eisenstein-Hu formulas and is included for completeness, however note that it is irrelevant once the growth factor is normalised to unity today. The reference wavenumber $k_\text{ref}$ is set to be the largest wavenumber in the table used by \texttt{class} that is less than $0.002$, and $z_\text{ref}$ is set to be the smallest redshift in the table used by \texttt{class} that is greater than 50.

\subsection{Comparison to WDM and FDM halo models}
\subsubsection{Theoretical comparison}
In \citet{BarkanaHaimanOstriker2001} the cutoff of the mass function at small masses for WDM was achieved by an additional mass dependence of the barrier (see also \citet{Marsh2016,BensonEtal2013}). This mass dependence of $\delta_c$ (a steep increase for small masses) was argued to be caused by the velocity dispersion, however it is unlikely that this is the true physical mechanism that suppresses the mass function below $M_{1/2}$ since WDM simulations have shown that the velocity dispersion is irrelevant for the large scale structure and the mass function \citep{SchneiderSmithReed2013,Vieletal2012}. This disparity was explained in \citet{SchneiderSmithReed2013} (p. 4 last paragraph before section 3) by splitting the effects of the velocity dispersion into two distinct time periods: the accumulated effect from times  $z>z_{\rm ini}$, (which manifests in the usual linear theory matter power spectrum cutoff), and the late time velocity dispersion (as should be present but turns out to be negligible in N-body simulations). 

In our GDM halo model, we have allowed for the possibility of both a cut-off of the matter power spectrum due to accumulated effects with $z>z_{\rm ini}$, \textit{and} a steepening of the barrier due to effects related to times $z<z_{\rm ini}$. Physically, $\delta^{\Lambda \rm GDM}_c(z , z_{\rm ini}, R)$, equation \eqref{eq_strongtimedep_DefGDM}, takes into account pressure and viscous effects that hinder collapse at $z>z_{\rm ini}$, whereas $\sigma^2_R(z_{\rm ini})$ is the integrated effect due to $z>z_{\rm ini}$. If $z_{\rm ini}$ is chosen during matter domination (as in our halo model) then only the latter (integrated) effect matters for WDM, and the cutoff in the WDM mass function must originate independent of late time velocity dispersion effects on $\delta_c$,  since it is observed in simulations without any added velocity dispersion as in \citet{SchneiderSmithMaccioEtal2012}. Thus the correct implementation of the mass function cut-off\footnote{In \citet{SchneiderSmithMaccioEtal2012} it was argued that this mass function cutoff is due to a non-hierarchical structure formation for masses below $M_{1/2}$ that is in conflict with the excursion set picture. However, it might be possible to show that this effect is the same as the strongly non-Markovian random walk, responsible for a mass function cut-off in \citet{ParanjapeLamSheth2012}, such that the cut-off can be understood within the excursion set theory.} due to non-Markovian behaviour caused by the linear theory power spectrum cutoff is not via the steep increase of the barrier for small masses when $z_{\rm ini}$ is chosen during matter domination. Instead it manifests through the phenomenological prefactor that is present in equation \eqref{eqn_fullgdmmultiplicity}, which depends on $M_{1/2}$.
{
Furthermore, when the WDM halo model is expressed in terms of our halo model, and $z_{\rm ini}$ is deep in the matter dominated era (as our $z_{\rm ini}=50$) then 
$$\delta^{\rm \Lambda WDM}_c(z , z_{\rm ini},R) \simeq \delta^{\rm \Lambda CDM}_c(z , z_{\rm ini})\,,$$ i.e. the spherical collapse threshold reduces to the scale-independent $\Lambda$CDM spherical collapse threshold. This is a special (approximate) property of all GDM models in which the GDM parameters grow strongly with redshift, like the WDM scaling $(1+z)^2$ of pressure and viscosity, such that the late universe dynamics is guaranteed to be more CDM-like compared to early times. In more general GDM models, and in particular for constant GDM parameters, we do not expect this approximation to hold, which is why we allow our halo model to have both a cut-off due to integrated earlier time behaviour and a later time change of the dynamics modelled by a modifed barrier.
}

Note that the mass dependence of $\delta_c^{\Lambda \rm GDM}$ we have introduced in section \ref{sec_gdmdeltacrit} for GDM, equation \eqref{eq_strongtimedep_DefGDM}, is due to the evolution of the shape of the linear power spectrum after $z_{\rm ini}$. There is (approximately) no shape change for WDM and thus our mass function is very similar to the one used in \citet{SchneiderSmithMaccioEtal2012}. The difference is that we allow for evolution of the shape of the power spectrum after $z_{\rm ini}$, which is expected for constant GDM parameters, and thus we have mass dependent $\delta_c^{\Lambda \rm GDM}$, which then requires using a mass function that can deal with mass dependent barriers.

\subsubsection{Qualitative numerical comparison}
In principle, we can compare our non-linear matter power spectrum to the spectra in literature (e.g. \citet{SmithMarkovic2011}, \citet{Vieletal2012}, \citet{SchneiderSmithMaccioEtal2012} and \citep{Marsh2016}. However, we note that we focus our work here on constant values of the GDM parameters, whereas WDM and FDM correspond to time and scale dependent parameters. Thus the impact of the non-linearities will be different, although we can nonetheless perform a qualitative comparison of how the halo model affects the predictions of our GDM model compared to how it affects the predictions for the specific cases of WDM and FDM. 

For all models, we see the same qualitative behaviour that the non-linear corrections increase the matter power spectrum and reduce the differences between $\Lambda$CDM and the modified matter content compared to the linear theory. However, there are differences in detail. For example, consider WDM with a mass of 0.25keV,  where meaningful changes to the linear spectrum compared to $\Lambda$CDM begin to occur on scales of $k\geq1h \text{Mpc}^{-1}$, and these changes begin on even smaller scales as the mass increases. For FDM, in line with \citep{Marsh2016}, the changes are on even smaller scales, similar to WDM with mass 1keV.  Whereas, at $k\geq1h \text{Mpc}^{-1}$, our GDM models consistent with Planck constraints differ from $\Lambda$CDM by two orders of magnitude in linear theory. This means that the GDM linear spectrum differs from $\Lambda$CDM on scales larger than those where the non-linear corrections matter for $\Lambda$CDM, whereas these two scales are swapped for the WDM and FDM models studied in the literature. This is because we have time-independent GDM parameters, whereas WDM would correspond to having them decay as $a^{-2}$, which causes their effects to appear only on small scales where nonlinearities are important. We expect that, once we allow for time dependent GDM parameters, the halo model will make a bigger difference relative to the linear spectrum, due to this reduced suppression in the linear theory at late times.

\section{Description of datasets and methodology}
\label{sec_method}
In this section we explain the data and methodology that were used to generate our results. We used the \texttt{class} code \citep{BlasLesgourguesTram2011}, modified as detailed in \citet{ThomasKoppSkordis2016,KoppSkordisThomas2016}, to evolve the GDM perturbation equations. We have added a module implementing the halo model as described in the previous section.

Our parameter constraints were obtained using the same basic methodology as in \citet{ThomasKoppSkordis2016}, see there for further details. We used the MCMC code MontePython \citep{AudrenLesgourguesBenabedetal2013} and established convergence of the chains using the Gelman-Rubin criterion~\citep{GelmanRubin1992}. We constrain a 6 parameter $\Lambda$CDM model  $\{\omega_b, \omega_g, H_0,n_s, \tau, \ln 10^{10} A_s\}$, where $\omega_g$ is the density of the dark matter fluid, which is CDM in the $\Lambda$CDM case and GDM otherwise. We set uniform priors on $\tau$ and $H_0$ such that $0.01<\tau$. The helium fraction was set to $Y_{\rm He}=0.24667$ \citep{PlanckCollaborationXIII2015} and we assumed adiabatic initial conditions. We used two massless and one massive neutrino with mass $0.06$ eV keeping the effective number of neutrinos to $N_\mathrm{eff} = 3.046$ (thus for simplicity we refer to ``neutrino mass'' during the analysis, although this is equivalent to the sum of the neutrino masses for this choice of parameters). The base parameter set is augmented by 3 GDM parameters $\{w,c^2_s,c^2_{\text{vis}} \}$ for the GDM runs, and additionally also the neutrino mass $m_\nu$ (for the single massive neutrino species) for some runs.

We perform runs for $\Lambda$CDM that are purely linear, linear+halofit and linear+halo model (where the halo model is as documented in the previous section, which is why it is important that our GDM halo model reduces to $\Lambda$CDM for vanishing GDM parameters). The \textit{halofit} \citep{SmithEtal2003,TakahashiEtal2012} runs are performed using the \textit{halofit} model built into \texttt{class}. For GDM, we will perform purely linear runs and linear+halo model runs, these runs will be referred to as ``HM'' in the results table.

Our primary dataset is the Planck 2015 data release \citep{PlanckCollaborationXI2015} of the CMB anisotropies power spectra, consisting of the low-$l$ T/E/B likelihood and the TT/TE/EE high-$l$ likelihood with the full ``not-lite'' set of nuisance parameters.\footnote{For full details, see the Planck papers and wiki http://wiki.cosmos.esa.int/planckpla2015/index.php/.}  These likelihoods combined are referred to as Planck Power Spectra (PPS).  We also use BAO\footnote{In appendix \ref{sec_continuityappendix} we examine a possible subtlety with the use of BAO data, which would also be relevant if GDM were constrained using redshift space distortion data.} from 6dF Galaxy Survey~\citep{BeutlerBlakeCollessEtAl2011} and the Baryon Oscillation Spectroscopic Survey Sloan Digital Sky Survey~\citep{AndersonAubourgBaileyEtal2014} (collectively referred to as BAO hereafter), and the Planck CMB lensing likelihood (Lens).

The key additional dataset that we use here is the WiggleZ matter power spectrum \citet{ParkinsonEtal2012} (referred to as MPS in the results table). This includes galaxy power spectrum measurements at four redshifts, $z=\{0.22, 0.41, 0.60, 0.78 \}$. We follow the procedure laid out in \citep{ParkinsonEtal2012} for the likelihood from this data, as implemented in MontePython, with the following exceptions. We do not use the WiggleZ giggleZ non-linear prescription, as this is not valid for GDM. Instead, we will use both \textit{halofit} and our halo model for $\Lambda$CDM runs, as detailed above. For $\Lambda$GDM, we will perform purely linear runs and runs using our halo model. Note that the WiggleZ likelihood processes the input theory spectrum, including convolving with the window function and other transformations. In particular, an analytic marginalisation over the linear bias is performed, see \citet{ParkinsonEtal2012} for details. We also consider two different subsets of the whole WiggleZ data: a conservative cut, where we use the provided k-bands up to $k=0.1 h$Mpc$^{-1}$, and a less-conservative cut using the complete data up to $k=0.3 h$Mpc$^{-1}$, the latter of which was used by the WiggleZ collaboration for their $\Lambda$CDM results.

\section{Constraints}
\label{sec_results}
We divide our constraints into two groupings: those without MPS data that focus on examining the robustness of previous constraints, and those using MPS data that aim to improve the constraints on GDM. The main results from these two groupings are that the previously obtained constraints are indeed robust and that the MPS data improves the constraints on $c^2_s$ and $c^2_\text{vis}$ by a factor of three. The constraints from the different runs can be found in table \ref{table_results}.

\subsection{Robustness of previous results}
\label{sec_robust}
Our first result follows from looking at table \ref{table_results}. Here we show our previous constraints on the GDM parameters using the data combinations PPS and PPS+Lens, both with and without the inclusion of the halo model (HM). It is clear that the constraints are essentially independent of the halo-model correction. This is because the halo model implemented as described above only has a small effect in the GDM matter power spectrum relative to the purely linear theory for the scales relevant to the upper limits of the constraints (see figures \ref{fig_lensingphi} and \ref{fig_spectra}). This is partly due to the choice of constant GDM parameters; as explained above we expect that the impact relative to the linear spectrum would be more important if we chose time dependent forms for the parameters, e.g. an $a^{-2}$ time dependence, or general binned functions. The relative effect of our halo model on the linear spectrum is decreased as the $c^2_s$ and $c^2_\text{vis}$ parameters are increased, as can be seen by comparing the GDM and $\Lambda$CDM halo model curves in figure \ref{fig_spectra}. This is caused by the strong effects on the linear spectrum from the $k_{\text{dec}}$ phenomenology, which dominate over any changes to the matter power spectrum due to the non-linear effects. Thus, we expect that any constraints on $c^2_s$ and $c^2_\text{vis}$ obtained from the CMB temperature, polarisation and lensing spectra on these scales  are actually more robust to potential non-linear complications than the standard $\Lambda$CDM parameters.

\begin{table*}
 \caption{Constraints on the GDM parameters for the two types of models and different combinations of experiments, for the $95\%$ and $99\%$ credible regions.}
\centering
\begin{mytabular}[1.8]{|l||cc|cc|cc||} 
 \hline 
 \hline 
Likelihood \hfill  Model  & \multicolumn{2}{|c|}{$10^2w$}  & \multicolumn{2}{|c|}{$10^6c_s^2$ (upper bound)}  & \multicolumn{2}{|c||}{$10^6c^2_\text{vis}$ (upper bound)}   \\
 (PPS+...) \hfill  ($\Lambda$-GDM+...) &  $95\%$ & $99\%$  &  $95\%$ & $99\%$ &  $95\%$ & $99\%$  \\
\hline 
 & $-0.040^{+0.473}_{-0.468}$  & $-0.040^{+0.700}_{-0.701}$   & $ 3.31$ & $ 6.31$ &  $ 5.70$ & $ 11.3$   \\
 + Lens &  $0.066^{+0.434}_{-0.427} $ & $0.066^{+0.654}_{-0.642}$ & $ 1.92$ & $ 3.44$ & $ 3.27$ & $ 5.99$  \\
 + Lens + BAO & $0.074^{+0.111}_{-0.110} $ & $0.074^{+0.164}_{-0.163}$ &  $ 1.91$ & $ 3.21$ &  $ 3.30$ & $ 6.06$  \\
 \hline 
 \hfill + HM & $-0.029^{+0.477}_{-0.481}$  & $-0.029^{+0.716}_{-0.690}$   & $ 3.11$ & $ 5.39$ &  $ 5.62$ & $ 11.1$   \\
 + Lens \hfill + HM & $-0.087^{+0.448}_{-0.461}$  & $-0.087^{+0.668}_{-0.649}$   & $ 1.92$ & $ 3.83$ &  $ 3.13$ & $ 5.79$   \\
\hline 
 + Lens + BAO \hfill  $+\ m_\nu$ & $0.101^{+0.159}_{-0.143}$  & $0.101^{+0.248}_{-0.201}$   & $ 1.90$ & $ 3.54$ &  $ 2.86$ & $ 4.82$   \\ 
\hline 
 + Lens + BAO + MPS ($k<0.1h \text{Mpc}^{-1}$)& $0.040^{+0.109}_{-0.108}$  & $0.040^{+0.164}_{-0.157}$   & $ 0.667$ & $ 1.21$ &  $ 1.10$ & $ 1.91$   \\
 + Lens + BAO + MPS ($k<0.1h \text{Mpc}^{-1}$)+ HM & $0.045^{+0.106}_{-0.109}$  & $0.045^{+0.161}_{-0.161}$  & $ 0.633$ & $ 1.11$ &  $ 0.953$ & $ 1.83$   \\
\hline 
 + Lens + BAO + MPS ($k<0.3h \text{Mpc}^{-1}$)& $0.035^{+0.112}_{-0.112}$  & $0.035^{+0.175}_{-0.168}$   & $ 0.0616$ & $ 0.103$ &  $ 0.0958$ & $ 0.16 $   \\
 + Lens + BAO + MPS ($k<0.3h \text{Mpc}^{-1}$)+ HM & $0.046^{+0.113}_{-0.111}$  & $0.046^{+0.169}_{-0.163}$   & $ 0.201$ & $ 0.254$ &  $ 0.333$ & $ 0.428$   \\
\hline 
 \hline 
 \end{mytabular} 
\label{table_results}
\end{table*}

\begin{figure*}
  \centering
   \includegraphics[width=6.in]{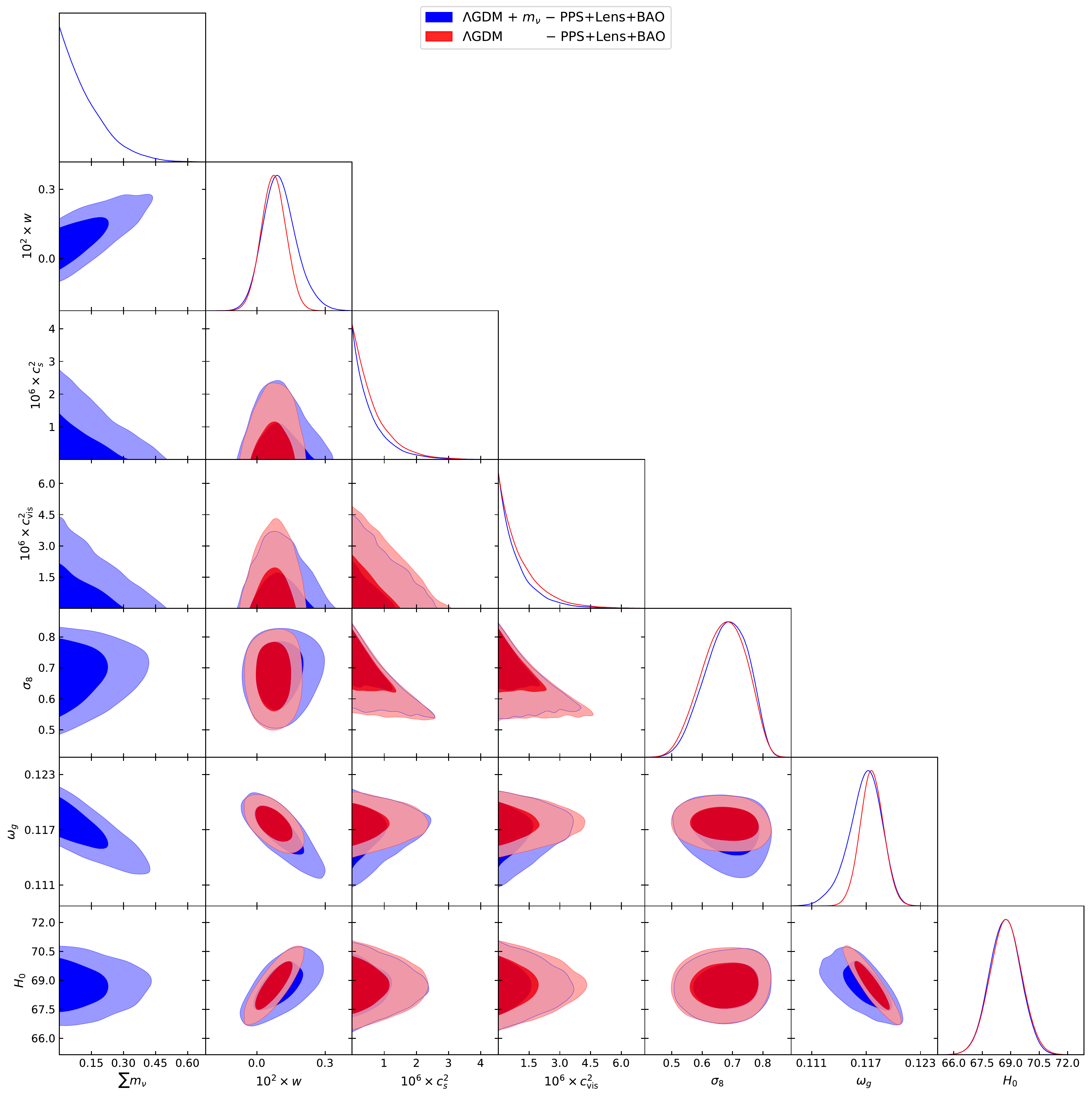}\hspace{-7.cm} \raisebox{8.9cm}{\includegraphics[width=3.3in]{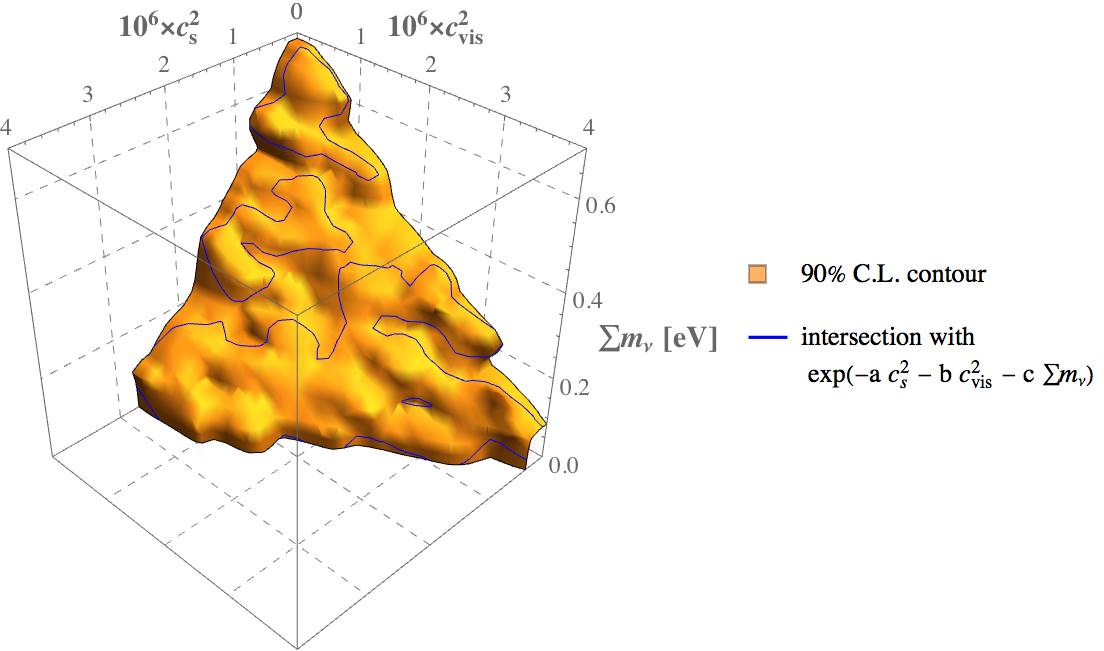}} \hspace{-5.9cm}\raisebox{4.8cm}{\includegraphics[width=2.6in]{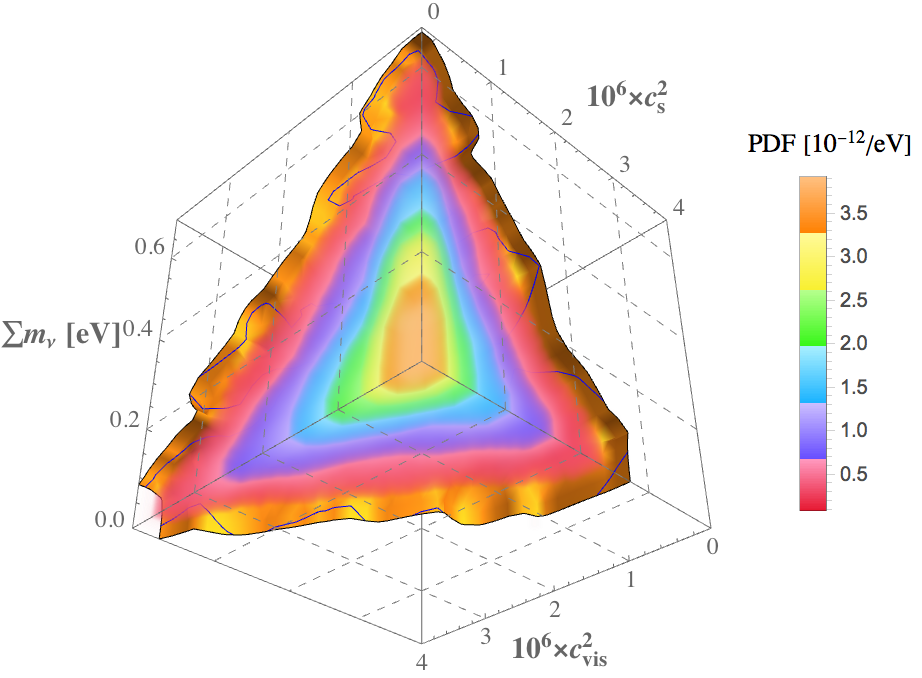}}
\caption{Posteriors for the neutrino mass and GDM parameters (plus other parameters of interest), where the red contours are for fixed neutrino mass and the blue contours are for when it is allowed to vary as an MCMC parameter. The 2D contours correspond to the $68\%$ and $95\%$ confidence levels. Changes to the 1D posteriors on the GDM parameters are visible when the neutrino mass is included as a parameter. This is due to the degeneracies that can be seen in the 2D posteriors: the neutrino mass is correlated with $w$ due to their similar impacts on the expansion history, and with the sound speeds due their similar impacts on CMB lensing. The right panels show the 3D posterior for $m_\nu, c_s^2$ and $c_\text{vis}^2$ which is peaked close to the origin and decreases further from this point, showing the expected degeneracy between all three parameters. For this posterior, the surfaces of constant confidence level are approximately planes, and are an extension of the $k_\text{dec}$ phenomenology found in \citet{ThomasKoppSkordis2016}, see section \ref{sec_robust} for details.
}
\label{fig_mnudegen}
\end{figure*}

Also in table \ref{table_results} we show the constraints for the data combination PPS+Lens+BAO, when the neutrino mass is both fixed or varied ($m_\nu$). The differences between the posteriors with and without the inclusion of the neutrino mass can be seen in figure \ref{fig_mnudegen}. The perturbative GDM parameters $c^2_s$ and $c^2_\text{vis}$ are affected little by the inclusion of the neutrino mass, indeed the constraints improve very slightly, whereas the inclusion of the neutrino mass does noticeably worsen the constraints on the equation of state $w$. These effects are both caused by the degeneracies between the neutrino mass and the GDM parameters, which can also be seen in figure \ref{fig_mnudegen}; we shall now explore these in more detail.

\begin{figure*}
  \centering
  \includegraphics[width=4.0in]{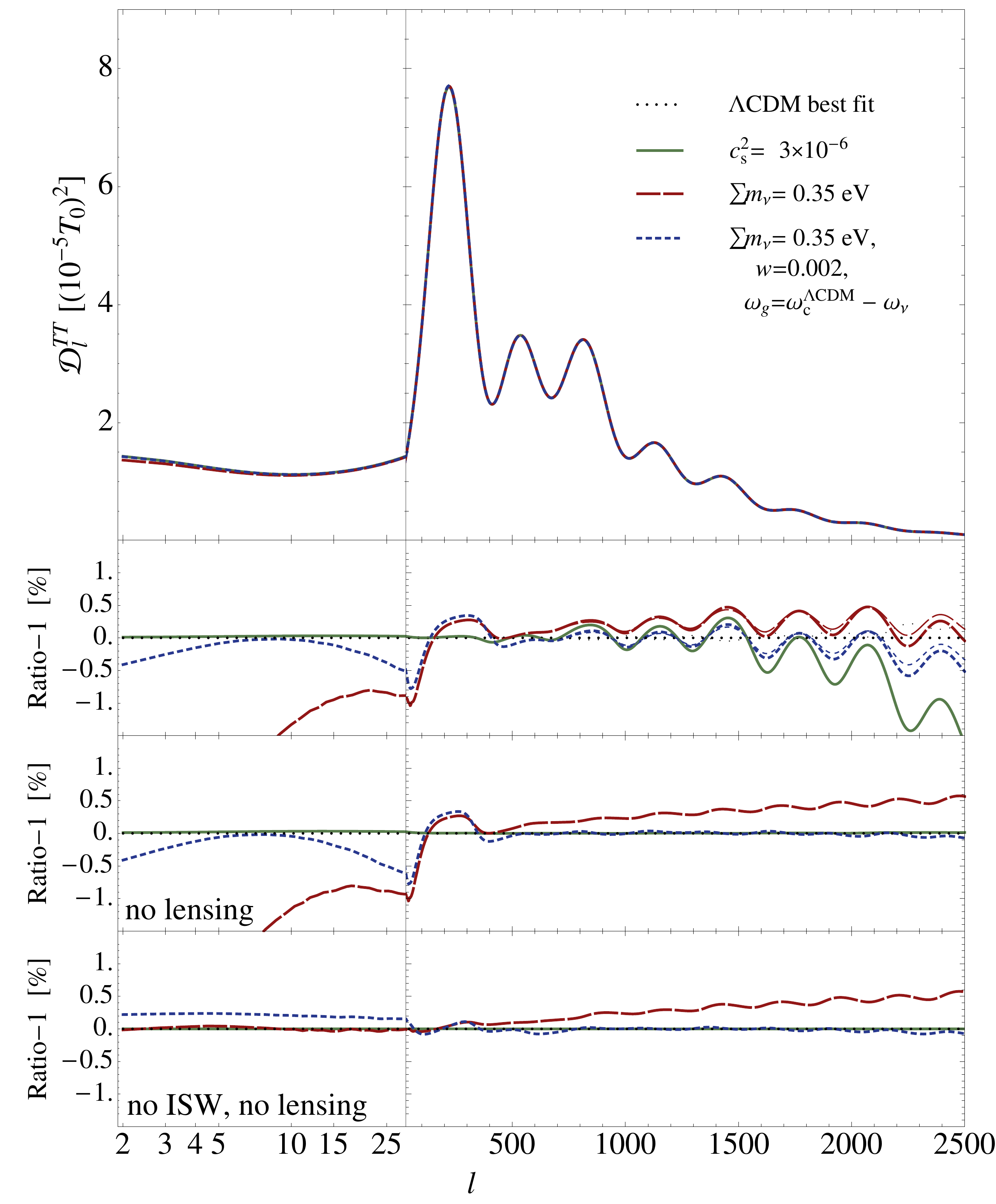}\begin{minipage}[b]{3.1in} 
   \includegraphics[width=2.9in]{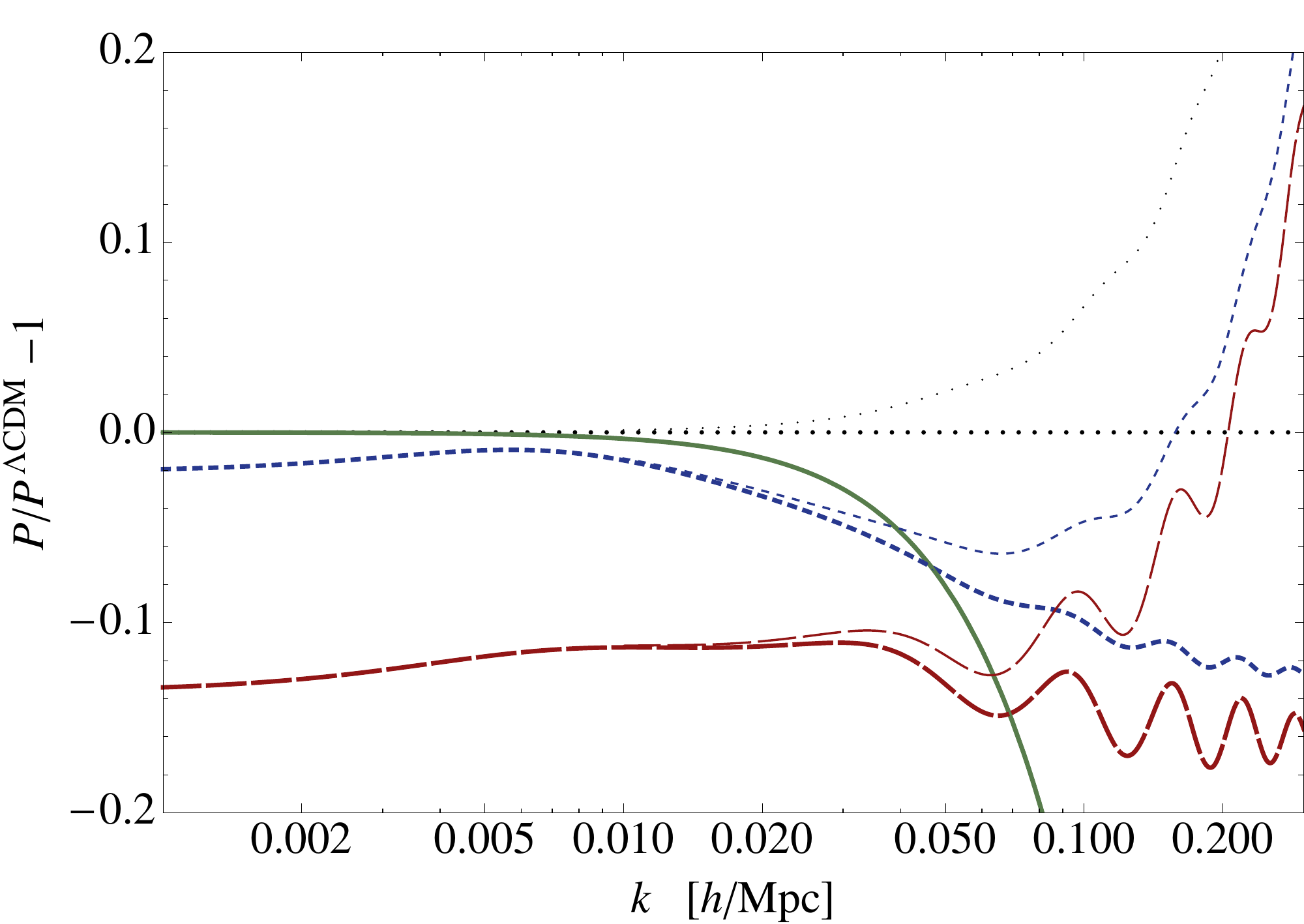} \\
  \vspace{1.0cm}
  \\
  \\
  \includegraphics[width=3.in]{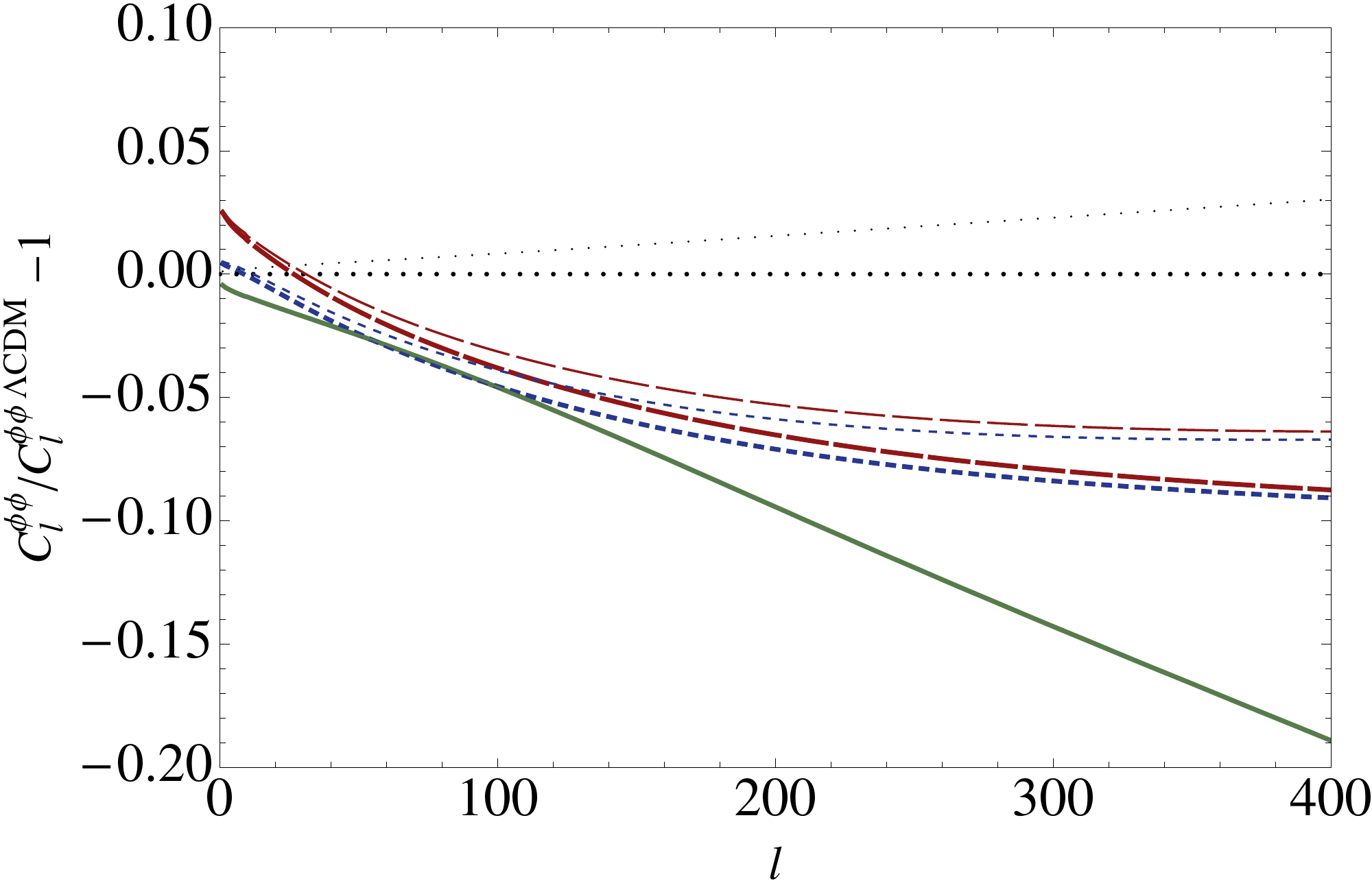}
  \end{minipage}
\caption{
On the left, we show the effect of GDM parameters and neutrino mass on the temperature power spectrum (left panel), the matter power spectrum (upper right panel) and the lensing potential (lower right panel). In all cases $\theta$ is kept fixed, so that the peak position remains at the same $l$ value. The three lower panels on the left show the ratio of the different spectra to the $\Lambda$CDM spectrum, where the lensing contribution has been removed from the second of these, and both the lensing and ISW contributions have been removed from the lowest panel. The full (green) and long-dashed (red) lines change the model by either turning on $c_s^2$ or increasing the neutrino mass from 0.06 eV to 0.35 eV, while keeping the DM abundance $\omega_g = \omega^{\rm \Lambda CDM}_c$ fixed.  Comparing the ratio of these models with the fiducial $\Lambda$CDM model in the second and third panels in the left plot shows why $c_s^2$ and $\sum m_\nu$ are degenerate: they reduce the amount of CMB lensing in a similar fashion. This is also clear by looking at the lower right plot displaying the lensing potential spectrum. The short dashed (blue) line maintains the increased neutrino mass and reduces the DM abundance, while at the same time increasing the DM equation of state from 0 to $w=0.002$. This shows why $w$ and $\sum m_\nu$ are degenerate: adjusting $w$ can make the expansion history of the massive neutrino cosmology more similar to the fiducial $\Lambda$CDM model, which can be seen by comparing the long dashed (red) line with the short-dashed (blue) line in third and fourth panel. The thinner version of these lines in the plots indicate the models that have been calculated using the halo model. The change of $C^{\phi \phi}_l$ for these models compared to $\Lambda$CDM is a direct consequence of the changes they cause for $P(k)$, see the upper right plot.
}
\label{fig_neumass}
\end{figure*}

The neutrino mass correlates with $c_s^2$ and $c_{\rm vis}^2$  in the same way that they are correlated with each other: The neutrino velocity dispersion is $c_\nu^2 = 2.78 \times 10^{-7}\, a^{-2} \left(1 \mathrm{eV}/m_\nu\right)^2$ \citep{LesgourguesPastor2006}, which causes a reduction in the lensing potential just like $c_s^2$ and $c_{\rm vis}^2$. This is not surprising since massive neutrinos can be described by a GDM fluid \citep{BlasLesgourguesTram2011}. This similarity between the effects of the neutrino mass and these GDM parameters can be seen by comparing the second and third panels of the left plot in figure \ref{fig_neumass}, showing the ratio of the spectra to $\Lambda$CDM with and without the lensing contribution. This is also shown in the lower plots of figure \ref{fig_neumass}, showing that all of these parameters result in a substantial reduction to the CMB lensing potential spectrum.  The right plots of figure \ref{fig_mnudegen} show the 3D posterior of $c_s^2$, $c^2_{\rm vis}$ and $\sum m_\nu$. The lower of the two plots is colour coded according to the probability density, which peaks in the $\Lambda$CDM corner, as expected due to the lack of a detection of GDM parameters and neutrinos mass. The upper of the two insets shows the 90\% confidence level contour in orange, which is shown to be well modelled by constant $c_s^2+0.6 c_{\rm vis}^2 + 3.9\times 10^{-6} \sum m_{\nu}[eV]$. This is in rough agreement with the expression for $c_\nu^2$, as expected if the degeneracy is indeed due to the reduction of the lensing potential described here.

Note that the geometry of this situation is slightly non-trivial: we are dealing with a ``corner'' of a multi-dimensional parameter space where the parameters are all required by physics to be non-negative. As noted in \citet{HeavensSellentin2018}, marginalising over parameters in such situations can cause some subtle effects, and this may be the source of the slight improvement to the constraints on $c_s^2$ and $c_{\rm vis}^2$ when the neutrino mass is included and marginalised over. 

Figure \ref{fig_mnudegen} also shows a clear degeneracy between $m_\nu$ and the equation of state $w$, which is due to the ability to generate cosmologies with identical $\theta$ but different $\omega_g$ when $w\neq0$. This can be seen by comparing the different panels in figure \ref{fig_neumass}, which show the CMB temperature power spectrum for different sets of parameters. Comparing the increased neutrino mass cosmology (red dashed) line in the third and fourth panels of the left plot, it can be seen that for fixed $\theta$ (the angular scale of the acoustic oscillations) and $\omega_c$ the main effect of the increased neutrino mass on a $\Lambda$CDM cosmology (aside from the lensing effect discussed above) is a reduction in the ISW effect and a tilt for higher $l$, see the ``no-lensing'' panel. The reduction of ISW (both early and late) as caused by the increased abundance non-relativistic matter content (compared to radiation and cosmological constant) when the neutrino mass is increased. One cannot simply compensate this with a change to $\omega_c$, because this would adversely affect expansion history at early times. However, when the parameter $w$ is introduced, it is possible to vary $\omega_g$, whilst also changing $w$ (and $H_0$) to keep the expansion history approximately fixed. This allows a GDM cosmology with increased neutrino mass to have the same ISW effect and high-$l$ tilt of $C^{TT}_l$ as a $\Lambda$CDM cosmology with lower neutrino mass, as can be seen by comparing the red (dashed) and blue (short-dashed) lines in the third panel on the left in figure \ref{fig_neumass}. This ability of $w, \omega_g$ to counteract these two effects of increasing $m_\nu$ drive the degeneracies between $m_\nu, w$ and $\omega_g$.

Note that the degeneracies between $m_\nu$ and the GDM parameters mean that if tighter constraints are put on the neutrino mass from other experiments, then using these results as a prior on CMB analysis could further improve the constraints on the GDM parameters.

\subsection{Use of MPS data}
\subsubsection{Conservative cut - No detection of GDM}
In the lower half of table \ref{table_results} we show the constraints obtained when including the WiggleZ matter power spectrum data (MPS), for two ranges of wavenumbers: $k<0.1h \text{Mpc}^{-1}$ and $k<0.3h \text{Mpc}^{-1}$. We discuss the former (more conservative) of these first. Note that for $\Lambda$CDM, the WiggleZ team found that the linear theory appears to give a better fit than \textit{halofit} (see figure 3 in \citet{ParkinsonEtal2012}), and we obtain the same result both for \textit{halofit} and the halo model, although the halo model has a smaller effect on the large scales than \textit{halofit}. We will focus the discussion on the perturbative parameters $c^2_s$ and $c^2_\text{vis}$, because the inclusion of the WiggleZ data has little effect on the constraints on $w$. This is because the effect of $w$ for the scales under consideration is primarily a small change to the amplitude, with little change to the shape of the spectrum, see \citet{KoppSkordisThomas2016}. Thus, the effect of $w$ will be almost entirely removed by the marginalisation over the bias in the WiggleZ likelihood, although note that future matter power spectrum data could constrain $w$ due to its effect on the peak location.

\begin{figure*}
  \centering
  \includegraphics[width=6in]{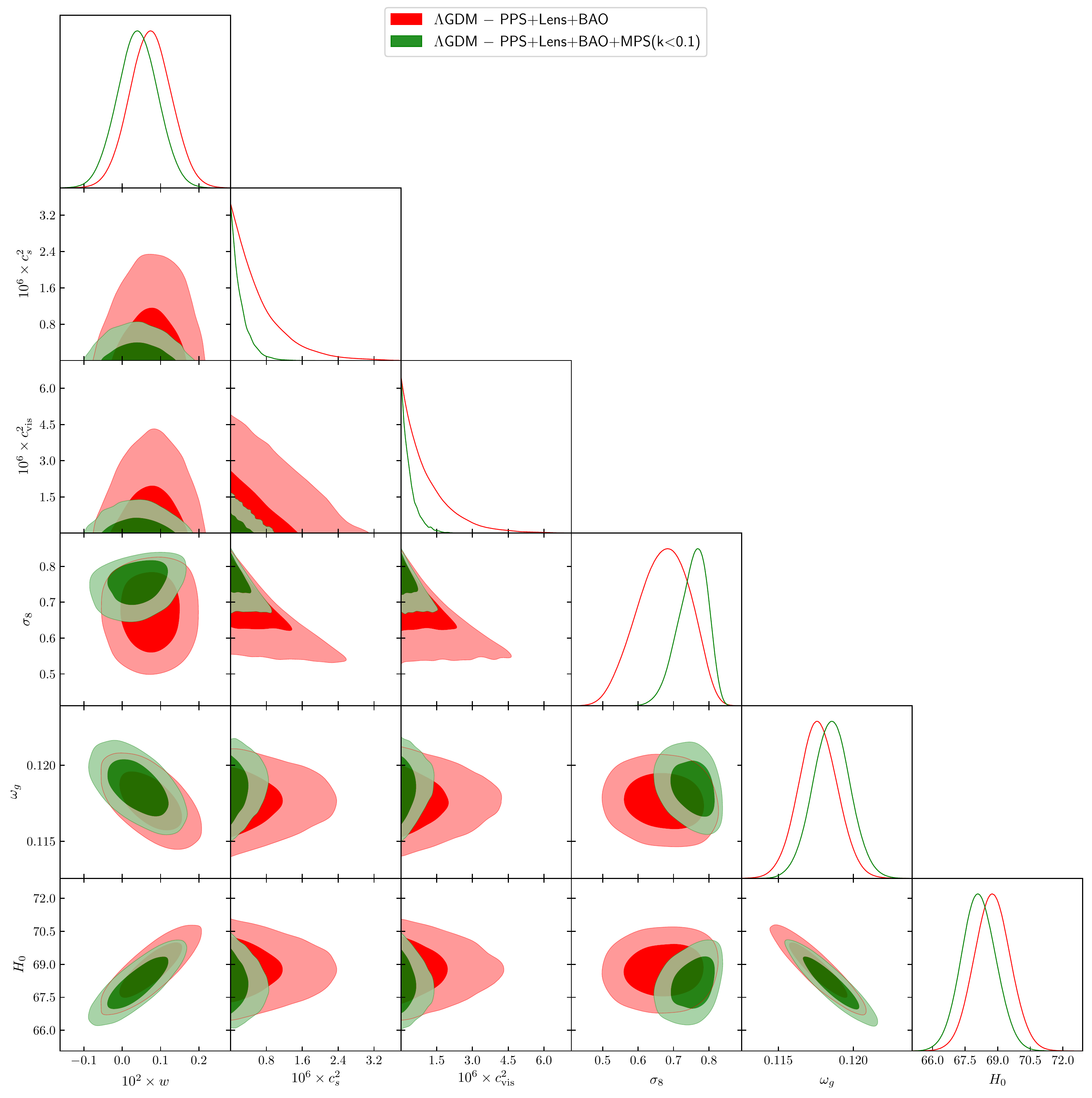}\hspace{-5.2cm}\raisebox{8cm}{\includegraphics[width=2.5in]{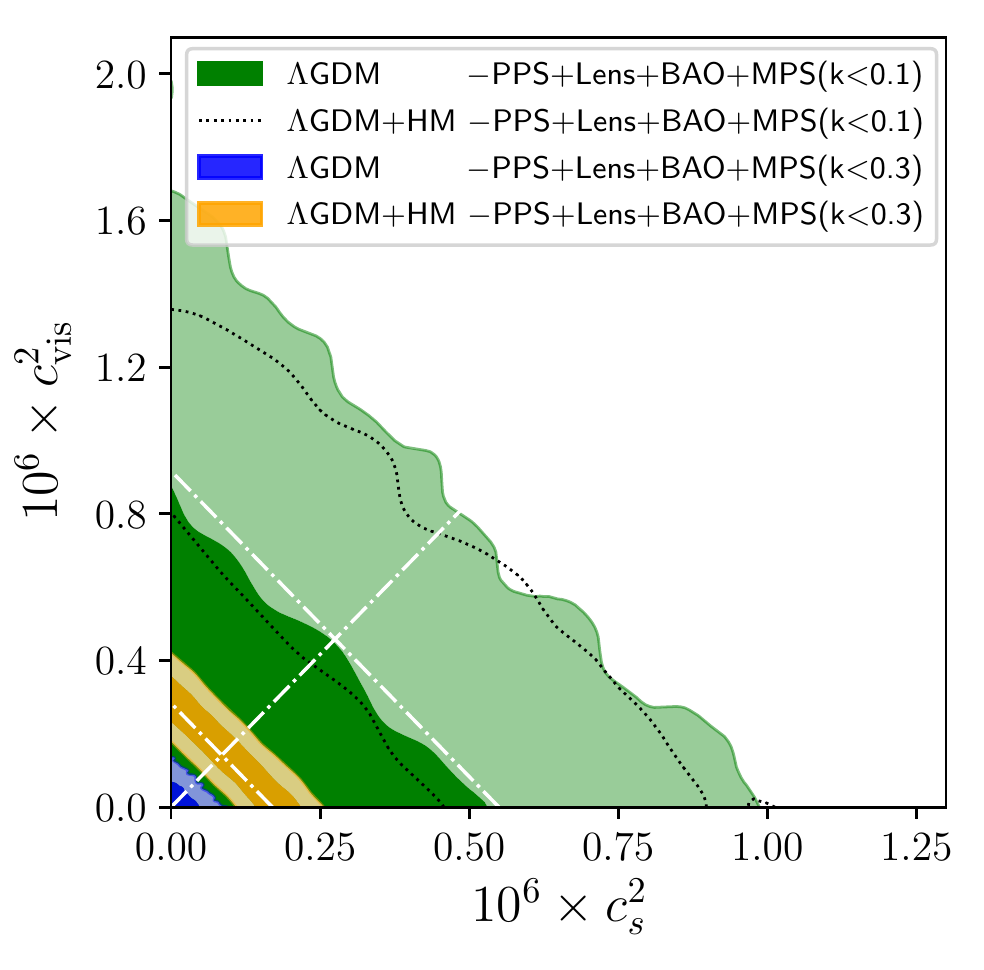}}
\caption{Posteriors of GDM and some cosmological parameters when WiggleZ data is used. The 2D contours correspond to the $68\%$ and $95\%$ confidence levels. The triangle compares our previous constraints \citep{ThomasKoppSkordis2016} (red) to those obtained when we include MPS data with a conservative cut $k<0.1h/$Mpc (green; these contours are for the linear implementation of GDM). The primary effect is a tightening of the 1D posteriors for $c_s^2$ and viscosity $c_{\rm vis}^2$. 
The right plot shows a more detailed comparison of constraints on these two parameters for the conservative and less-conservative cuts, for both linear and halo model modelling of the GDM matter power spectrum. Here, the green (filled) contours and black (dotted line) contours show the constraints obtained for the conservative cut, for linear and halo model GDM respectively. These contours show that  including quasi-linear scales $k<0.1 h/$Mpc is robust: the constraints are not sensitive to the inclusion of the halo model. The two smaller sets of contours show the constraints for the less-conservative cut ($k<0.3 h/$Mpc): the blue contours are for linear modelling of GDM and the yellow contours are for the constraints obtained with the halo model. The GDM parameters are now more strongly constrained for both sets of less-conservative contours, however the halo model shows a clear preference for $\Lambda$GDM over $\Lambda$CDM ($c_s^2=c_{\rm vis}^2=0$), see the yellow contours, while there is no such preference if we use the linear theory to fit the data (the blue contours). This is an indication that we currently cannot robustly constrain GDM parameters using these smaller scales and that more work needs to be done. The white dashed lines indicate the direction of constant $c_s^2 + 0.6 c_{\rm vis}^2$ following the $k_\text{dec}$ phenomenology and the direction perpendicular to this, which is the most strongly constrained direction. 
}
\label{fig_cs2andcv2HM}
\end{figure*}

The inclusion of the matter power spectrum with the conservative cut has a strong effect on the perturbative parameters $c^2_s$ and $c^2_\text{vis}$, improving the constraints by a factor of three, see table \ref{table_results}. This can also be seen in figure \ref{fig_cs2andcv2HM}, where the 1D posteriors narrow considerably when the WiggleZ data is included. As this figure shows, this improvement is the only significant change to the posteriors due to the extra data. The green contours in this figure are from the linear implementation of GDM, however we note that the changes due to the extra data are approximately independent of whether the linear or halo model implementation of GDM is used; this can be seen both from the constraints in table \ref{table_results} and in the right panel of figure \ref{fig_cs2andcv2HM} by comparing the green (linear) and black dotted (halo model) contours. This is important because it implies that the constraints using the conservative cut come from physics that is well understood, and is not sensitive to detailed considerations of the non-linear regime.

\begin{figure}
  \centering
 \includegraphics[width=\columnwidth]{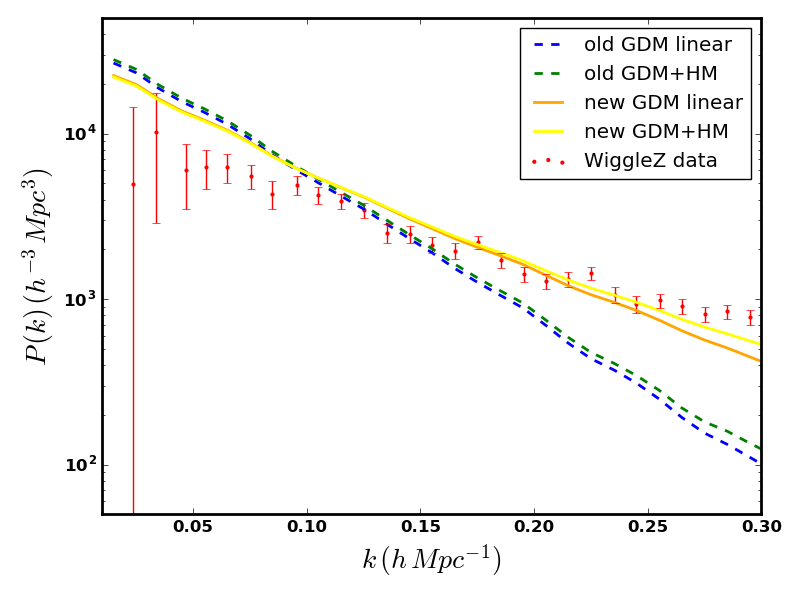}\\
 \includegraphics[width=\columnwidth]{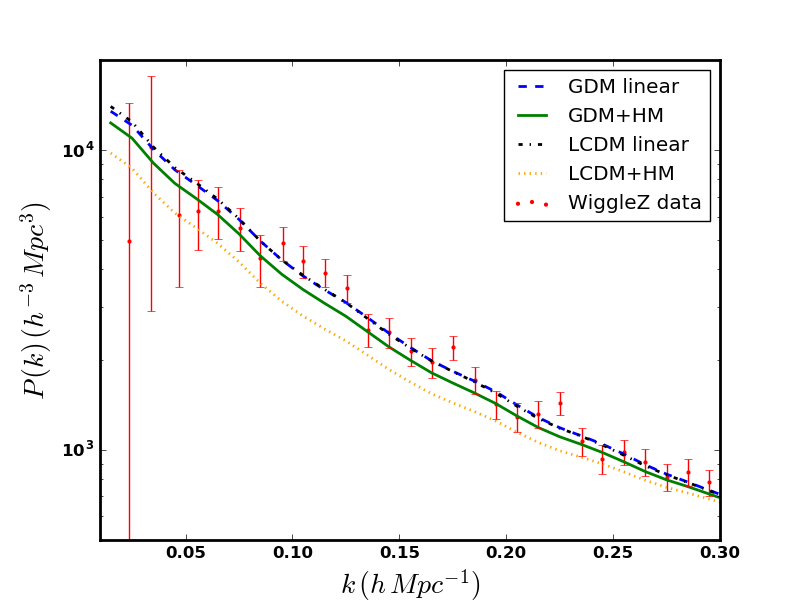}
\caption{Comparison of GDM and $\Lambda$CDM theoretical spectra to WiggleZ data for the lowest redshift bin; In both plots, the red points show the data.
\textit{Upper:} GDM curves with $c^2_\text{vis}=0$ and non-zero $c^2_s$ (but note that the plot for $c^2_s=0$ and non-zero $c^2_\text{vis}$ is essentially identical, since the degeneracies between these parameters hasn't been broken). The (dashed) blue and green curves correspond to the linear and halo-model predictions respectively, for parameters corresponding to our previous (linear) constraints in \citet{ThomasKoppSkordis2016} ($c^2_s=0.000003$). The (solid) orange and yellow curves correspond to the linear and halo-model predictions respectively, with GDM parameters corresponding to the improved constraints in this paper when the MPS data is used with the conservative cut ($c^2_s=0.000001$). It can be seen that the previous constraints have some tension with the WiggleZ data, and thus that its inclusion improves the constraints on the GDM parameters by requiring smaller values of the parameters to reduce the tension. For all parameter choices here, the difference between the linear and non-linear spectra is small.
\textit{Lower:} Theoretical spectra constructed using the best fit parameters from the MCMC runs with the less-conservative cut for the WiggleZ data. The spectra correspond to linear $\Lambda$CDM (black dot-dashed), $\Lambda$CDM with the halo model (orange dotted), linear GDM (blue dashed) and GDM with the halo model (green solid). The linear GDM best fit parameters are $c^2_s=6.284\times10^{-10}$ and $c^2_\text{vis}=2.55\times10^{-8}$ and the GDM plus halofit best fit parameters are $c^2_s=1.6\times10^{-7}$ and $c^2_\text{vis}=4.3\times10^{-8}$. Similarly to WiggleZ \citep{ParkinsonEtal2012}, we get a better fit for linear $\Lambda$CDM than when the halo model is included. The best fit model for linear GDM has small GDM parameters and a spectrum that is very similar to the $\Lambda$CDM spectrum, as expected since the constraints in this case are consistent with $\Lambda$CDM. For the GDM halo model case, the best fit model has a much larger sound speed, and thus deviates more from the $\Lambda$CDM best fit.
}
\label{fig_mpscs2halomodel}
\end{figure}

The improvement on the $c^2_s$ and $c^2_\text{vis}$ constraints is due to the decay of the matter power spectrum for $k>k_{\rm dec}$, which creates a slope to the matter power spectrum that is inconsistent with the data for larger values of $c^2_s$ and $c^2_\text{vis}$. This can be seen in the upper panel of figure \ref{fig_mpscs2halomodel}, which compares theoretical spectra with various upper limits of GDM parameters to the lowest redshift bin in the WiggleZ data. Note that in both panels of this figure the theory spectra are transformed in line with how the likelihood is computed, in order to be compared to the data. This includes convolving with the survey window function, marginalising over the linear bias and accounting for the difference in background to the fiducial cosmology. See \citet{ParkinsonEtal2012} for a full description and explanation of these processes. The upper panel of figure \ref{fig_mpscs2halomodel} compares GDM spectra computed with non-zero $c^2_s$ values corresponding to upper limits of previous constraints (``old'') and the constraints when the matter power spectrum data is included (``new''). There is a tension between the slope of the theoretical spectra and the data that is reduced when the lower value (associated to the constraints from the WiggleZ data) is used. The equivalent plot for non-zero $c^2_\text{vis}$ is essentially identical, since the degeneracy between these parameters has not been broken (see below).  In accordance with what was noted earlier, the use of the halo model makes little difference to the theoretical spectra in figure \ref{fig_mpscs2halomodel}, and thus to the constraints obtained with and without the halo model. The constraining power here does not come from the measured amplitude of the matter power spectrum because of the marginalisation over the linear bias. We expect that these constraints could increase even further if the matter power spectrum is measured on larger scales, particularly the turnover around the peak. This would have the additional advantage of staying inside our conservative regime that we have seen is robust to non-linear modelling. Considering how the difference in slopes shown in figure \ref{fig_mpscs2halomodel} continues for $k>0.1h \text{Mpc}^{-1}$, we expect that the less-conservative cut for the WiggleZ data will improve the constraints further, see below.

\begin{figure}
  \centering
 \includegraphics[width=\columnwidth]{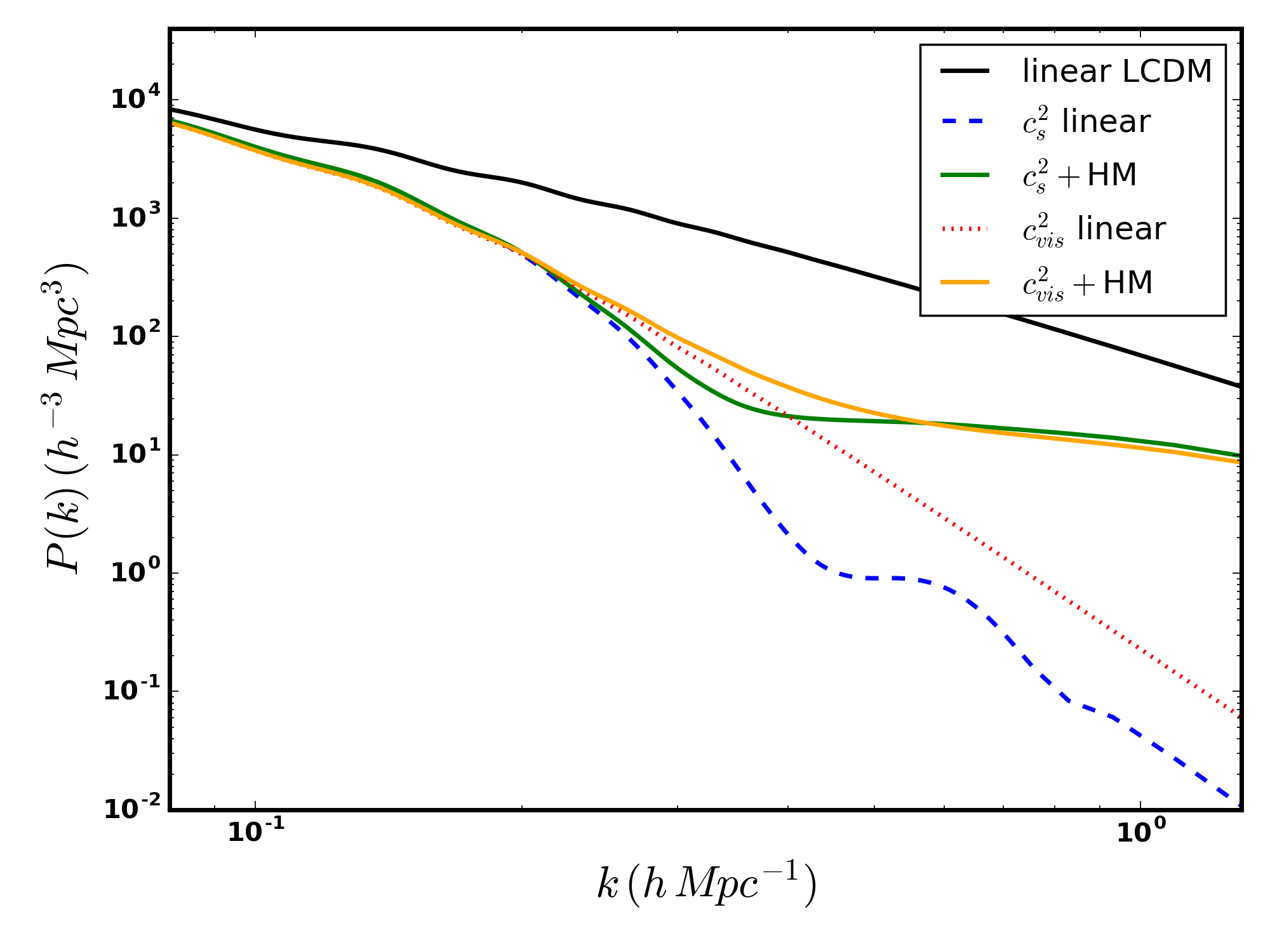}
\caption{The linear and non-linear matter power spectra for the two perturbative GDM parameters. The black curve shows the linear $\Lambda$CDM spectrum. The blue (dashed) and green (solid) curves show the linear and non-linear spectra for non-zero $c^2_s$ and the red (dotted) and orange (solid) curves show the linear and non-linear spectra for non-zero $c^2_\text{vis}$. The values of $c^2_s$ and $c^2_\text{vis}$ are chosen to produce the same value of $k_\text{dec}$. Up to the level of our conservative cut ($k=0.1h \text{Mpc}^{-1}$), there is little difference between the two linear spectra and between the two non-linear spectra, and this difference only begins to manifest close to the smallest scales in our less-conservative cut ($k=0.3h \text{Mpc}^{-1}$). Note that on larger scales, the non-linear modelling acts to recreate the degeneracy between $c^2_s$ and $c^2_\text{vis}$.
}
\label{fig_gdmspectra}
\end{figure}

One of the motivations for considering matter clustering data for GDM constraints is the attempt to break the degeneracy between $c^2_s$ and $c^2_\text{vis}$ caused by the $k_\text{dec}$ phenomenology. The scales we are looking at with the WiggleZ data here are insufficient to break this degeneracy, see figure \ref{fig_gdmspectra}: up to $k=0.1h \text{Mpc}^{-1}$, there is little difference between $c^2_s$ and $c^2_\text{vis}$. Note that from \citet{KoppSkordisThomas2016} we expect the value where the oscillations start to be $k_J = 1/(0.2 c_s \tau)\simeq 0.2$, which is in agreement with what we see here. Even on scales down to $k=0.3h \text{Mpc}^{-1}$ (i.e. to the level of our less-conservative cut; see below), the difference between the spectra generated by the two parameters is not large, although beyond $k=0.3h \text{Mpc}^{-1}$  the difference between the linear spectra increases substantially. Interestingly, the non-linear spectra corrections act to recreate the degeneracy between the $c^2_s$ and $c^2_\text{vis}$ spectra on scales below approximately $k=0.5h \text{Mpc}^{-1}$ (see figure \ref{fig_gdmspectra}). If this modelling of GDM non-linearities is accurate then this means that there is only a small range of scales (around $k=0.3h \text{Mpc}^{-1}$), in which data could allow us to distinguish the effects of these two parameters.

\subsubsection{Non-conservative cut - Possible detection of GDM}
As mentioned above, we also consider a less conservative cut, with $k<0.3h \text{Mpc}^{-1}$. The constraints are presented in the final two lines of table \ref{table_results}, where it can be seen that a significant gulf opens between the constraints obtained with and without the halo model. The constraints including the halo model improve by a factor of 3 compared to the more conservative cut, amounting to a combined improvement compared to previous constraints of an order of magnitude. However, the constraints without the halo model are another factor of 3 or so stronger still.  This weakening of the constraints due to the inclusion of the halo model naively seems to match our expectations of weaker results when the non-linear effects are included, however there is a deeper story here.

The right panel of figure \ref{fig_cs2andcv2HM} shows the 2D posterior contour plots for  $c^2_s$ and $c^2_\text{vis}$ for the conservative and less-conservative cuts, for both linear and halo model implementations of GDM. As discussed above, the linear and halo model versions of GDM result in very similar constraints for the conservative cut to the matter power spectrum data. The less-conservative cut for linear GDM results in a similar looking set of contours, in the sense that the contours are all essentially right-angled triangles with a similar slope on the side joining the two axes. I.e. the contours are solely upper bounds with a particular slope and the two GDM parameters both being zero is consistent with the data. The only difference is that the upper bounds have been significantly reduced.

However, the shape of the less-conservative cut for GDM with the halo model is significantly different; the lower contour is now apparent, resulting in a trapezoidal contour and a clear inconsistency of the $\Lambda$CDM point ($c^2_s=0$ and $c^2_\text{vis}=0$) with the contours. There is now a clear preference for a non-zero GDM parameter. This difference between the linear and halo model GDM results is further seen by looking at the right panel of figure \ref{fig_mpscs2halomodel}. Here we plot the spectra generated using the best fit parameters from the different MCMC runs. The linear GDM best fit parameters are $c^2_s=6.284\times10^{-10}$ and $c^2_\text{vis}=2.55\times10^{-8}$ and the GDM plus halofit best fit parameters are $c^2_s=1.6\times10^{-7}$ and $c^2_\text{vis}=4.3\times10^{-8}$. The best fit model for linear GDM has small GDM parameters and a spectrum that is very similar to the $\Lambda$CDM spectrum, as expected since the constraints in this case are consistent with $\Lambda$CDM. For the GDM halo model case, the best fit model has a much larger sound speed, and thus deviates more from the $\Lambda$CDM best fit. Thus we can see the discrepancy between the linear GDM and halo model GDM manifesting here as well. Interestingly, the GDM halo model run returns a best fit spectrum that is closer to the linear $\Lambda$CDM spectrum than the halo model $\Lambda$CDM spectrum is. The $\chi^2$ for the two GDM curves and the linear $\Lambda$CDM curve are almost indistinguishable,\footnote{We should caution here that an MCMC code such as MontePython is not optimised in terms of finding the lowest possible value of the likelihood, so we expect that the best fit values we have found here are close to the absolute minimum, but not precisely the lowest values.} showing further that the inclusion of the halo model into GDM can compensate for higher values of the GDM parameters.

Despite the preference for a non-zero GDM parameter shown in the GDM+HM contours in the right panel of figure \ref{fig_cs2andcv2HM}, the individual 1D posteriors (not shown here as their interpretation is dubious; see below) show no preference to be non-zero due to the degeneracy between the two parameters. This means that marginalising over the other parameter results in no ``detection'' of a non-zero value of either parameter. Despite the resulting large difference in upper bounds between the less-conservative results with and without the halo model, the maximal width of the contours along the $45^\circ$ line marked in the plot is similar in the two cases.

The slope of the degeneracy in these contours is well understood as the direction along which $k_\text{dec}$ remains fixed, and it is the direction perpendicular to this (the $45^\circ$ line marked in the plot) that is most constrained by the data. In principle, it would be possible to create a new parameter that describes this direction (i.e. essentially $c^2_+ \equiv c^2_s+8/15\, c^2_\text{vis}$), and then the second coordinate (i.e. $d\equiv (1 + 8/15\, c_{\rm vis}^2/c_s^2)^{-1}$, $0\leq d \leq 1$) of this 2D space can be marginalised over, in order to create a 1D posterior for $c^2_+ $.  If we had chosen uniform priors on $c^2_+$ and $d$, we would expect the 1D-posterior of this new $c^2_+ $-parameter to peak at non-zero values for the yellow contour in Fig.\ref{fig_cs2andcv2HM}, and to peak at zero for the blue contour. And thus this parameter would allow to compactly quantify the detection of perturbative GDM parameters. However, we note that our current choice of priors (uniform priors on $0\leq c_s^2<0.1$ and $0\leq c_{\rm vis}^2<0.1$) would make $c^2_+$ peak at non-zero value for any sensible 2D posterior, in particular also for the blue contour in Fig.\ref{fig_cs2andcv2HM}. We will explore these issues in a forthcoming paper.

We note that this difference between the linear and non-linear results for the less-conservative cut is a strong justification of the motivation behind this paper, namely that correctly modelling these non-linear scales will be crucial for using late-time clustering data to constrain the GDM parameters. At the level of the modelling we have done here, we cannot be sure of our results with either the linear or non-linear modelling. Instead, we take these results to show that there is a need to look into the non-linear modelling in substantially more detail, before a detection of non-zero GDM parameters using late time clustering data could be claimed. Furthermore, even if the less-conservative linear contours agreed with the non-linear contours, then we would be hesitant to claim a detection of non-zero GDM parameters because of the caveats related to marginalising over parameter subspaces with special geometries. Nonetheless, our results show that late time matter clustering data can strongly constrain the GDM parameters, to the level where a detection is possible.

\section{Conclusion}
\label{sec_conc}
We have investigated how considerations around large scale structure can affect constraints on the generalised dark matter parameters. The main results of this work are the development of the halo model for GDM presented in section \ref{sec_halo} and the improved constraints on the GDM parameters presented in table \ref{table_results}.

In section \ref{sec_halo} we argued for modifying the $\Lambda$CDM halo model in a particular way, by backtracking the ``linearly-extrapolated'' critical density for collapse. This allows the mass dependence of the collapse barrier to be implemented in a natural way. This halo model reduces to a standard $\Lambda$CDM halo model in the case of scale independent growth, and produces qualitatively similar results to \textit{halofit}. Having derived the halo model for GDM, we note that the non-linear corrections are much less significant for GDM (with constant $c_s^2$ and $c_{\rm vis}^2$) than for $\Lambda$CDM, because the strong linear decay  dominates over the corrections from the halo model. 

We use this halo model to test the robustness of previously obtained constraints based on Planck CMB power spectra, as seems circumspect considering the magnitude of the $\Lambda$CDM non-linear corrections and difference between $\Lambda$CDM and GDM spectra in figure \ref{fig_lensingphi}. We find that the GDM constraints change little, as expected from the aforementioned phenomenology of the GDM halo model. Interestingly, the perturbative GDM parameters $c^2_s$ and $c^2_\text{vis}$ are less sensitive to the non-linear corrections than the standard $\Lambda$CDM parameters, which will be increasingly important for future CMB lensing surveys, such as Simons Observatory \citep{Simons2018}. We additionally checked the changes to previous GDM constraints when the neutrino mass is allowed to vary as a free parameter, primarily finding a worsening of the constraints on the equation of state (of GDM) $w$, due to the degenerate effects on the expansion history. We also elucidate the three-way degeneracy between $m_\nu$, $c^2_s$ and $c^2_\text{vis}$, and note that the geometry of this situation requires that marginalisation over these parameters is done carefully \citep{HeavensSellentin2018}.

We examined the effect of including the WiggleZ matter power spectrum data when constraining the GDM parameters, finding a factor of three improvement on the sound speed $c^2_s$ and viscosity $c^2_\text{vis}$ constraints when a conservative cut in wavenumber $k$ is used. When increasing the $k$-range that is included, these constraints improve by a further factor of three, for a total improvement from the use of matter power spectrum data of an order of magnitude. This shows the value of datasets that constrain the $k_\text{dec}$ phenomenology of GDM. Since we analytically marginalise over the linear bias, we expect that once large scale structure measurements reach the peak of the matter power spectrum, the total constraining power will be sufficient to either constrain the GDM parameters to the point of cosmological irrelevance or yield a detection of beyond $\Lambda$CDM physics.\footnote{Note that, being more precise, this ``beyond $\Lambda$CDM physics'' could just be a more careful and precise modelling of the $\Lambda$CDM universe, according to the EFTofLSS interpretation of GDM.} These improved constraints are one of the key results of this work.

The results from extending the $k$-range that is included show some important features for future work. The first is that there is a difference between the constraints obtained by linear and non-linear modelling, thus showing the importance of robust non-linear modelling of the GDM model for the use of late time matter clustering data. This suggests that a re-evaluation of the tight constraints obtained in \citet{KunzNesserisSawicki2016} could be interesting. This also shows that future surveys, combined with a more detailed analysis of the non-linear completion of GDM, have the potential to show that the matter power spectrum is not consistent with the $\Lambda$CDM. Furthermore, we have shown that GDM with the halo model yields a 2D contour that is clearly inconsistent with the $\Lambda$CDM point ($c^2_s=c^2_\text{vis}=0$), although we advise caution in the interpretation of this due to the difference between the linear and non-linear results. Even in this case, neither $c^2_s$ nor $c^2_\text{vis}$ is individually detected, due to the degeneracy between these two parameters, and we have noted that it is not straightforward to quantify a detection in such cases, because marginalisation in areas of parameter space with corners can lead to biases. Doing so requires a careful analysis of priors, see e.g. \citet{HeavensSellentin2018}. 
We plan to explore this issue further in the future. 

As the $\gamma_1,\gamma_2$ parameters in the halo model were originally calibrated for WDM, it may be preferable to treat them as nuisance parameters and vary them in an MCMC analysis for GDM. However, this further increases the computational demands of the already expensive codes and is unlikely to make a difference to our results as we have made no detection of GDM. Furthermore many other aspects of the halo model, like the precise form of the spherical collapse barrier, remain currently an educated guess and require a detailed study using numerical simulations of a suitably defined non-linear GDM model. Thus we leave an investigation into these issues for future work.

Throughout this work we used the simplest parameterisation of the GDM parameters; a single value with no redshift or scale dependence. We expect the halo model to have a larger impact for time dependent GDM parameters, although we leave this investigation to future work. In particular, we expect that for $c^2 \propto a^{-2}$, corresponding to WDM and FDM, that the halo model has a larger impact. { This is because the late universe GDM parameters will have a much smaller impact on the DM dynamics for fixed early universe values, so that the dynamics can be approximated by CDM in the late universe, allowing non-linearities to develop despite the small scale decay of the linear matter power spectrum imprinted during early times}. The assumption of constant GDM parameters was relaxed in \citet{KoppEtal2018}, where the equation of state $w$ was measured in multiple redshift bins. The importance of the WiggleZ data considered here for the constraints on $c^2_s$ and $c^2_\text{vis}$ suggest that this data could be crucial for putting strong constraints on redshift and scale dependent forms for $c^2_s$ and $c^2_\text{vis}$. Given the results here for $w$ when including $m_\nu$, it would be interesting to revisit the results using time dependent $w$ bins from \citet{KoppEtal2018}. In particular, it may be the case that the neutrino mass is only degenerate with $w$ over certain redshift ranges, and thus the time variation of $w$ allows the degeneracy with $m_\nu$ to be broken, and most of the constraining power on $w$ to be recovered. In addition we note that if the GDM parameters are given specific time (and scale) dependence corresponding to either FDM or WDM, then it would be interesting to perform an in-depth quantitative comparison of the existing FDM and WDM halo models with the halo model presented here for general GDM models. That comparison is beyond the scope of this paper.

\section*{Acknowledgements}
The research leading to these results has received funding from the European Research Council under the European Union's Seventh Framework Programme (FP7/2007-2013) / ERC Grant Agreement n. 617656 ``Theories
 and Models of the Dark Sector: Dark Matter, Dark Energy and Gravity''. The Primary Investigator is C. Skordis. DM acknowledges support from the UK Science \& Technology Facilities Council through grant ST/N000668/1 and from the UK Space Agency through grant ST/N00180X/1. DBT acknowledges support from Science and Technology Facilities Council (STFC) grant ST/P000649/1. MCMC chains for this analysis were partially run on the Sciama High Performance Compute (HPC) cluster which is supported by the ICG, SEPNet and the University of Portsmouth. We thank the developers of  \texttt{class}, \texttt{montepython} and \texttt{getdist} for making their codes public. We thank the WiggleZ collaboration for making their data and likelihood public, and thank D. Parkinson for help with understanding these. We thank C. Skordis and S. Ili\'c for helpful discussions. This research has made use of NASA's Astrophysics Data System.

\bibliographystyle{mnras}
\bibliography{gdmhalomodel} 

\appendix

\section{$\Lambda$CDM halo model}
\label{sec_LCDMhalo}
\begin{figure}
  \centering
 \includegraphics[width=\columnwidth]{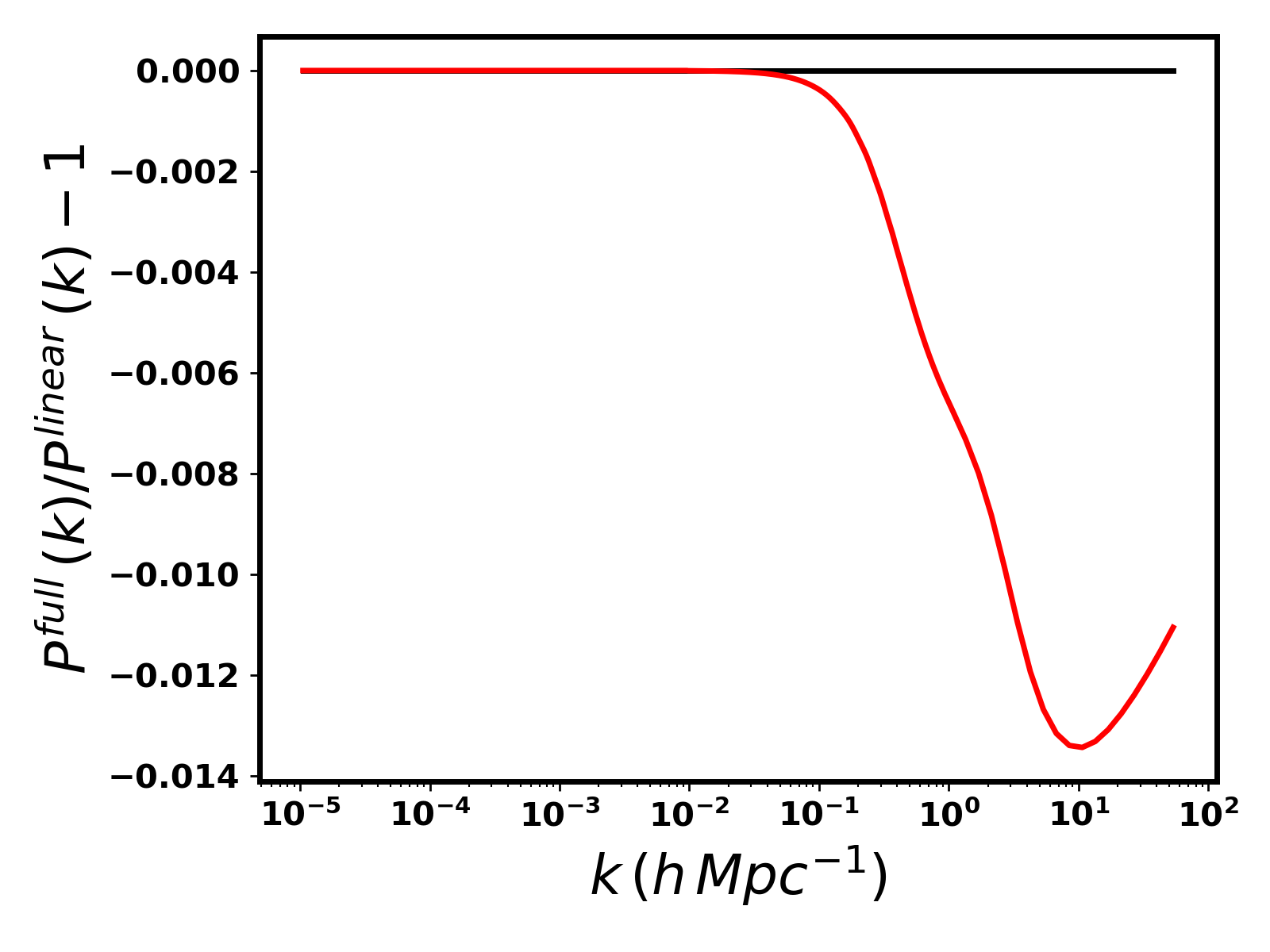}
\caption{Fractional change to the $\Lambda$CDM matter power spectrum when the full 2-halo term is used, rather than when it is approximated by the linear spectrum. The effect of including the full 2-halo term is less than $0.5\%$ on the scales of interest.
}
\label{fig_full2haloratio}
\end{figure}

Here we present the standard $\Lambda$CDM halo model, concepts and notation in detail, allowing us to focus on the development of the GDM halo model in the main text. The summary here draws on \citet{CooraySheth2002,MeadPeacockHeymansEtal2015,Marsh2016}. As stated in the text, the halo model is a semi-analytic method for computing the matter power spectrum on non-linear scales. It starts by writing $\rho_m(\mathbf{x}) = \sum_i \rho_{\rm halo}(\mathbf{x} - \mathbf{x}_i)$  with $\langle \rho_m(\mathbf{x}) \rangle = \int dM M \frac{dn}{dM} = \bar{\rho}_m$ which gives the matter power spectrum, the Fourier transform of $\langle (\rho_m(\mathbf{x})-\bar \rho_m) (\rho_m(\mathbf{x+r})-\bar\rho_m) \rangle$, as a sum of ``1-halo'' and ``2-halo'' terms,
\begin{equation}
P_m(k)=P_{1h}(k)+P_{2h}(k)\text{,}
\end{equation}
where the 1-halo term corresponds to correlations within halos, and the 2-halo term corresponds to correlations between halos.  Thus, the 2-halo term typically dominates on large scales and the 1-halo term typically dominates on small scales.  As the density field is described as a superposition of (spherically symmetric) halos, for randomly distributed halos, the power spectrum would thus have the form of shot noise (that takes into account the internal structure of the halos). This is described by the 1-halo term
\begin{equation}
P_{1h}(k)=\frac{1}{\bar{\rho}_m^2}\int^\infty_0 M^2 \frac{dn}{dM} \tilde{u}^2(k,M) dM\text{,}
\end{equation}
where $\bar{\rho}_m=3\Omega_{m0} H^2_0/8\pi G$ is the background matter density in the Universe and $\Omega_{m0}$ is the background density today in units of the critical density. The integral is over the mass $M$, $\frac{dn}{dM}$ is the mass function and $\tilde{u}(k,M)$ is the normalised Fourier transform of the halo density profile $\rho_{\rm halo}(|\mathbf{x}|)$. The 2-halo term is given by
\begin{equation}
P_{2h}(k)=\left[\frac{1}{\bar{\rho}_m}\int^{M_\text{max}}_{M_\text{min}}dM M \tilde{u}(k,M)\frac{dn}{dM} b_1(M) \right]^2 P_\text{lin}(k)\text{,}
\end{equation}
where $b_1(M)$ is the linear bias.The 2-halo term tends towards the linear spectrum $P_\text{lin}(k)$ on large scales due to the consistency relation
\begin{equation}
\frac{1}{\bar{\rho}_m}\int^{\infty}_{0} dM M \frac{dn}{dM} b_1(M)=1\text{.}
\end{equation}
Indeed, in our code we set the 2-halo term to be the linear spectrum because the loss of accuracy is very small (see figure \ref{fig_full2haloratio}, where we show the change in the $\Lambda$CDM spectrum as a result of this approximation) and reduces the computational time required for each call to the halo model code.
We will now look more closely at the ingredients for the 1-halo term.

\subsection{Mass function}
There are many mass functions in the literature, ranging from the original Press-Schechter formulation \citep{PressSchechter1974}, to the Sheth-Tormen model \citep{ShethTormen1999} based on elipsoidal collapse and models calibrated against up-to-date N-body simulations. Those various mass functions differ in their explicit functional form of the so-called multiplicity function $f(\sigma)$, from which the mass function can be calculated as
\begin{equation} \label{massfunc}
\frac{dn}{dM}dM=\frac{\bar{\rho}_m}{M}f(\sigma)\frac{d \ln \sigma^{-1}}{dM}dM \text{,}
\end{equation}
where $dn$ is the comoving halo number density and $\sigma$ is the dimensionless matter power spectrum smoothed on a scale $R$,
\begin{equation} \label{sigma}
\sigma_R^2(z)=\frac{1}{2\pi^2}\int^\infty_0 P_\text{lin}(k,z)W^2(kR) k^2 dk \text{.}
\end{equation}
The scale $R$ is related to the enclosed mass through
\begin{equation}
M= \frac{4 \pi }{3} R^3 \bar{\rho}_m \,,
\end{equation}
such that $M$, $R$ and $\sigma$ can all be used interchangeably. A typical smoothing function is the spherical (real space) top hat,
\begin{equation}
W^2(x)=\frac{3}{x^3}\left( \sin x -x \cos x \right) \text{.}
\end{equation}
We will use the mass function of Achitouv, Corasaniti, Maggiore and Riotto (ACMR)\citep{CorasanitiAchitouv2011,MaggioreRiotto2010}, as this has been proven useful in other extensions of $\Lambda$CDM \citep{KoppApplebyAchitouvEtal2013, AchitouvWagnerWellerEtal2014, AchitouvBaldiPuchweinEtal2015}, and is  physically well motivated. $f(\sigma)$ takes the form \eqref{ftot}
\begin{align} \label{MarkovianAchizcollapse}
f(\sigma)&=\frac{\bar B- \sigma^2 d \bar B/ d\sigma^2}{\sigma}\sqrt{\frac{2 a_b}{\pi}}e^{-\frac{a_b}{2\sigma^2}\bar B^2} + ...\text{,} \\
\bar B &= \delta_\text{crit} + \beta \sigma^2\,, \label{asphericalBarierAchizcollapse}
\end{align}
where the parameters take the values $\beta=0.12$ and $a_b=0.7143$, and the barrier $\bar B$ is composed of the spherical collapse barrier $\delta_\text{crit}$ and mass-dependent correction $\beta \sigma^2$. 
This latter takes into account that small halos (large $\sigma$) are more difficult to form (and therefore have higher $\bar B$) due to more likely asphericity. 
The ellipses stand for non-Markovian corrections of the mass function that arise through correlations of the density field at different smoothing scales, see equation \eqref{ftot}. 
For CDM $\bar B- \sigma^2 d \bar B/ d\sigma^2=\delta_\text{crit}$. The critical ``linearly-extrapolated'' (see section \ref{sec_halo}) density for spherical collapse, $\delta_\text{crit}\approx1.686$ in an Einstein-deSitter universe. For a $\Lambda$CDM universe, we use a form with a mild redshift and cosmology dependence from \citet{NakamuraSuto1997}
\begin{equation}
\delta_\text{crit}=\frac{3}{20}\left(12\pi \right)^\frac{2}{3}\left(1+0.012299\log \Omega_m(z) \right) \text{.}
\end{equation}

\begin{figure}
  \centering
 \includegraphics[width=\columnwidth]{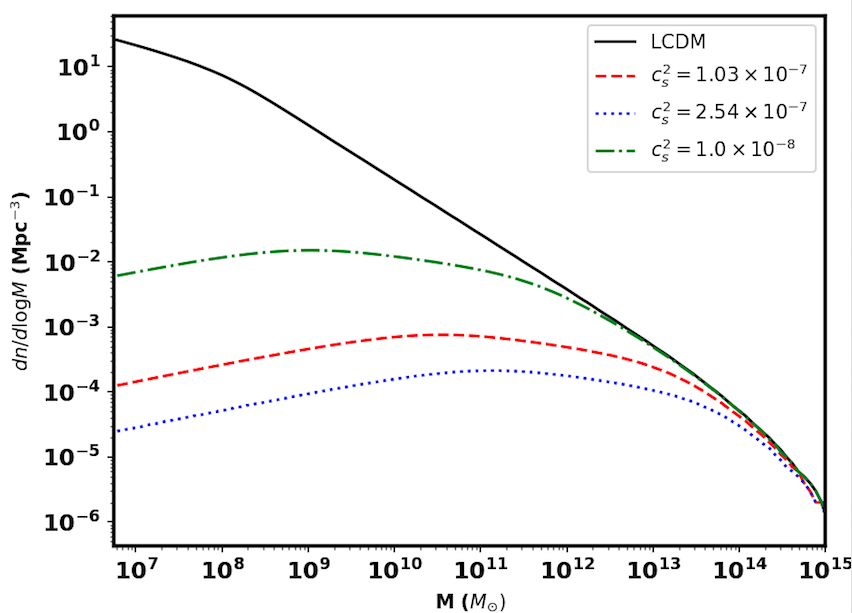}
\caption{The halo mass function as used in our halo model, for $\Lambda$CDM and 3 different GDM parameter choices, for redshift $z=0$.
}
\label{fig_hmf_z0}
\end{figure}

{In figure \ref{fig_hmf_z0}, we show the full halo mass function as developed in section \ref{sec_halo}, for $\Lambda$CDM and several choices for the GDM parameter $c^2_s$. We note that constant values of the GDM parameters, as used throughout this paper, have a strong effect on the mass function, since the pressure and viscosity have a continuous effect during collapse in addition to the initial cutoff in the power spectrum. This is in contrast to GDM parameters that have a time dependence, such as the $a^{-2}$ dependence exhibited by the GDM parameters for WDM, where only the initial cutoff affects the mass function. Due to the speculative nature of our mass function, we have focussed on the matter power spectrum as a more robust observable; we leave it to future work to investigate the mass function and halo model in greater detail, and consider to what extent our constraints can be improved by e.g. galaxy cluster counts once the mass function has been validated in greater detail.}

\subsection{Halo density profile}
The Fourier transform of the halo density profile is given by
\begin{equation}
\tilde{u}(k,M)=\frac{1}{M}\int^{r_v}_0 dr \frac{\sin (kr)}{kr} 4\pi r^2 \rho_\text{halo}(r,M)\text{.}
\end{equation}
The most commonly used halo density profile is the Navarro-Frenk-White (NFW) profile \citep{NavarroFrenkWhite1996}
\begin{equation}
\rho_\text{NFW}=\frac{\rho_0}{\left(r/r_s \right)\left(1+r/r_s\right)^2} \text{,}
\end{equation}
where $\rho_0$ is a normalisation constant. The virial radius $r_v$ defines the extent of the halo, $M=\frac{4}{3}\pi \bar{\rho} \Delta_v r^3_v$, where $\Delta_v$ denotes the overdensity of the virialised halo with respect to the average background density in the universe. For a $\Lambda$CDM universe, we can use the approximate relation, again from \citet{NakamuraSuto1997},
\begin{equation}
\Delta_v=18\pi^2 \Omega_m^{-0.573}  \text{.}
\end{equation}
The scale radius $r_s$ is typically related to the concentration $c\equiv r_v/r_s$. There are several parameterisations used for the concentration, we will use one that is used in \citet{BullockKolattSigadEtal2001,PielorzRodigerTerenoEtal2010},
\begin{equation}
c(M,z)=\frac{c_\star}{1+z}\left( \frac{M}{M_\star}\right)^{-\alpha}\text{,}
\end{equation}
with $c_\star=10$ and $\alpha=0.2$.  The parameter $M_\star$ is the solution to
\begin{equation}
\sigma\left(z=0,M_\star \right)=\delta_\text{crit}(z=0)\text{.}
\end{equation}

The Fourier transform of the NFW profile can be expressed as 
\begin{eqnarray}
\tilde{u}=\left[\ln(1+c) -\frac{c}{1+c} \right]^{-1}\left(\sin \eta \left[Si (\eta[1+c]) -Si(\eta)\right] \right.\nonumber\\
\left.+\cos \eta \left[Ci(\eta[1+c])-Ci(\eta) \right]-\frac{\sin (c\eta)}{(1+c)\eta}\right)\text{,}
\end{eqnarray}
where $\eta=kr_\text{v}/c$ and we have defined
\begin{eqnarray}
Ci(x)&=&-\int^\infty_x \frac{\cos t}{t}dt\\
Si(x)&=&\int^x_0 \frac{\sin t}{t}dt \text{.}
\end{eqnarray}

When calculating the 2-halo term precisely, the bias for the Markovian part of the ACMR mass function is given by
\begin{equation}
b_1(M)=1 + a_b \beta - \frac{1}{\delta_\text{crit}} + \frac{a_b \delta_\text{crit}}{\sigma^2}  \text{,}
\end{equation}
where the parameters take the same values as in the mass function above \citep{CorasanitiAchitouv2011,AchitouvBaldiPuchweinEtal2015}.

\subsection{Comment on large scale power}
\label{sec_halomodellargescale}
The 1-halo term contributes a constant power on large scales, which has the effect of increasing the matter power spectrum from the halo model above that of the linear theory, on scales where the density contrast is small and therefore the linear theory would be expected to apply. This effect is due to the un-compensated nature of the NFW profile, meaning that only positive density perturbations are included. Thus, there is a shot-noise-like term added by the 1-halo term on large scales, that is unphysical if the dark matter power spectrum is the desired output. On the other hand, a finite sample of galaxies would have an intrinsic shot-noise component, which could conceivably dominate over the linear power on very large scales. In principle, this problem could be remedied by using a compensated halo profile, however this results in the halo model having zero power on large scales. A good discussion of this issue is in \citet{CooraySheth2002}.

This issue is considered further in \citet{SeljakVlah2015}. Following that work, we implement a compensation term for the 1-halo part of the halo model. Accordingly, the 1-halo term is multiplied by a compensation function $F(k)$
\begin{equation}
F(k)=1-\frac{1}{1+k^2R^2}\text{,}
\end{equation}
where the compensation scale $R$ is given by
\begin{equation}
R=\frac{26\,\mathrm{Mpc}}{h}\left( \frac{\sigma_8(z)}{0.8}\right)^{0.15}=\frac{26 \,\mathrm{Mpc}}{h}\left( \frac{\sigma_8(z=0)}{0.8}\frac{D(z)}{D(z=0)}\right)^{0.15}\text{,}
\end{equation}
and $D(z)$ is the growth factor. Note that $F(k)\rightarrow0$ as $k\rightarrow0$, meaning that the 1-halo term vanishes as the 2-halo term tends towards the linear spectrum. Thus recovering the expected large scale result.

\subsection{Some relevant details of the numerical implentation}
Here we just note a couple of details of the numerical implementation fo our halmodel in the \texttt{class} code for those who wish to duplicate our results. Note that the $Si$ and $Ci$ integrals above have numerical solutions, see e.g. \citet{MeadPeacockHeymansEtal2015} or Wikipedia. Also note that it is important to set the \texttt{class} parameter ``P\_k\_max\_1/Mpc'' to be sufficiently large; we use the value 30.0.

When sampling the mass function we work with a number of logarithmically spaced masses, where the upper limit of the mass range is given by $10^{16} M_\odot$ and the lower mass limit is set to $M_\text{min}=\log\left(\frac{4.0\pi \bar{\rho}}{3.0 k^3_\text{min}}\right)\approx 10^8 M_\odot$. The number of masses was set to 150, but we note that we checked our results for convergence against these three parameters and our results are robust to sensible variations in these parameters.

\begin{figure}
  \centering
 \includegraphics[width=\columnwidth]{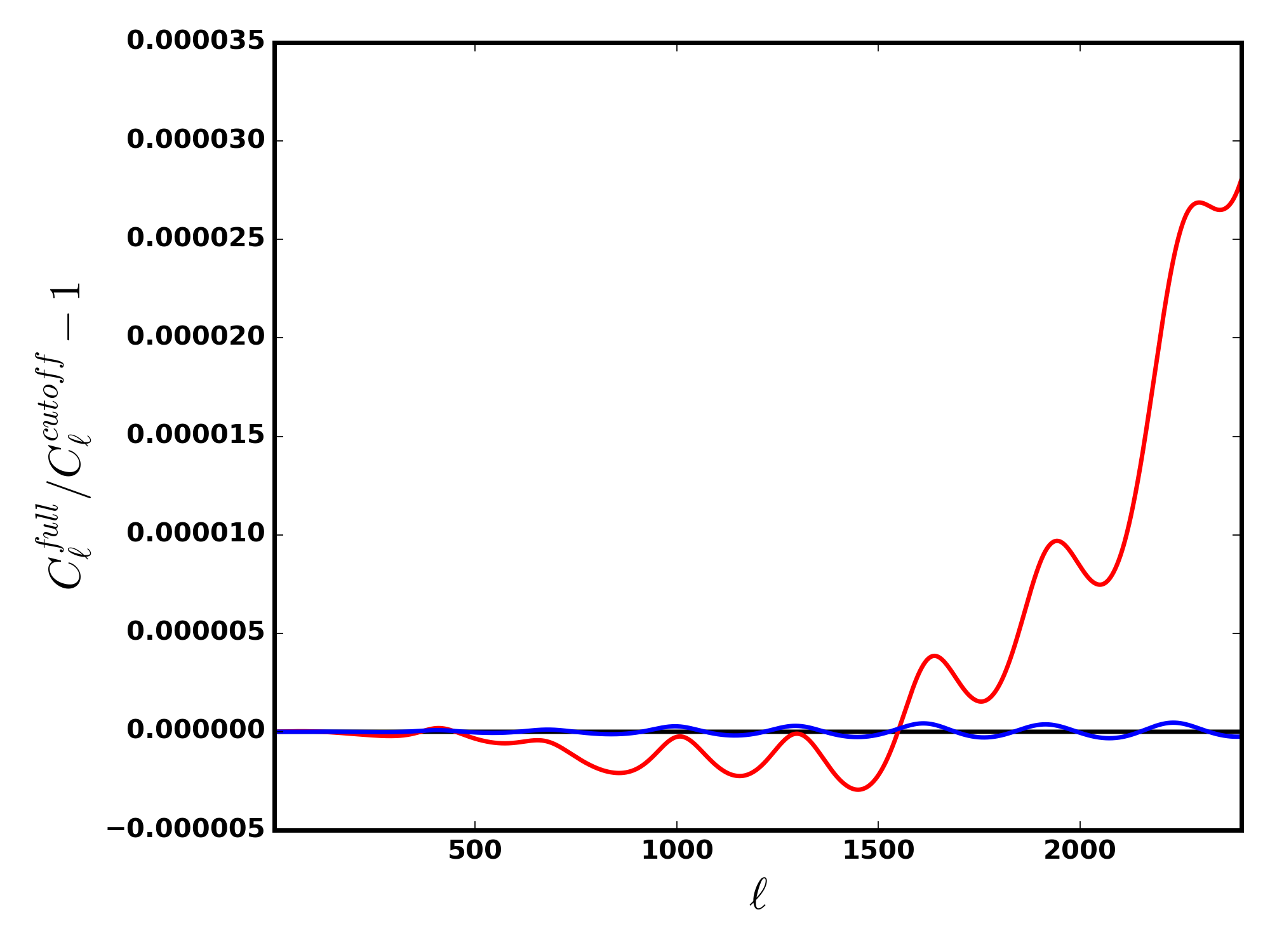}
  \includegraphics[width=\columnwidth]{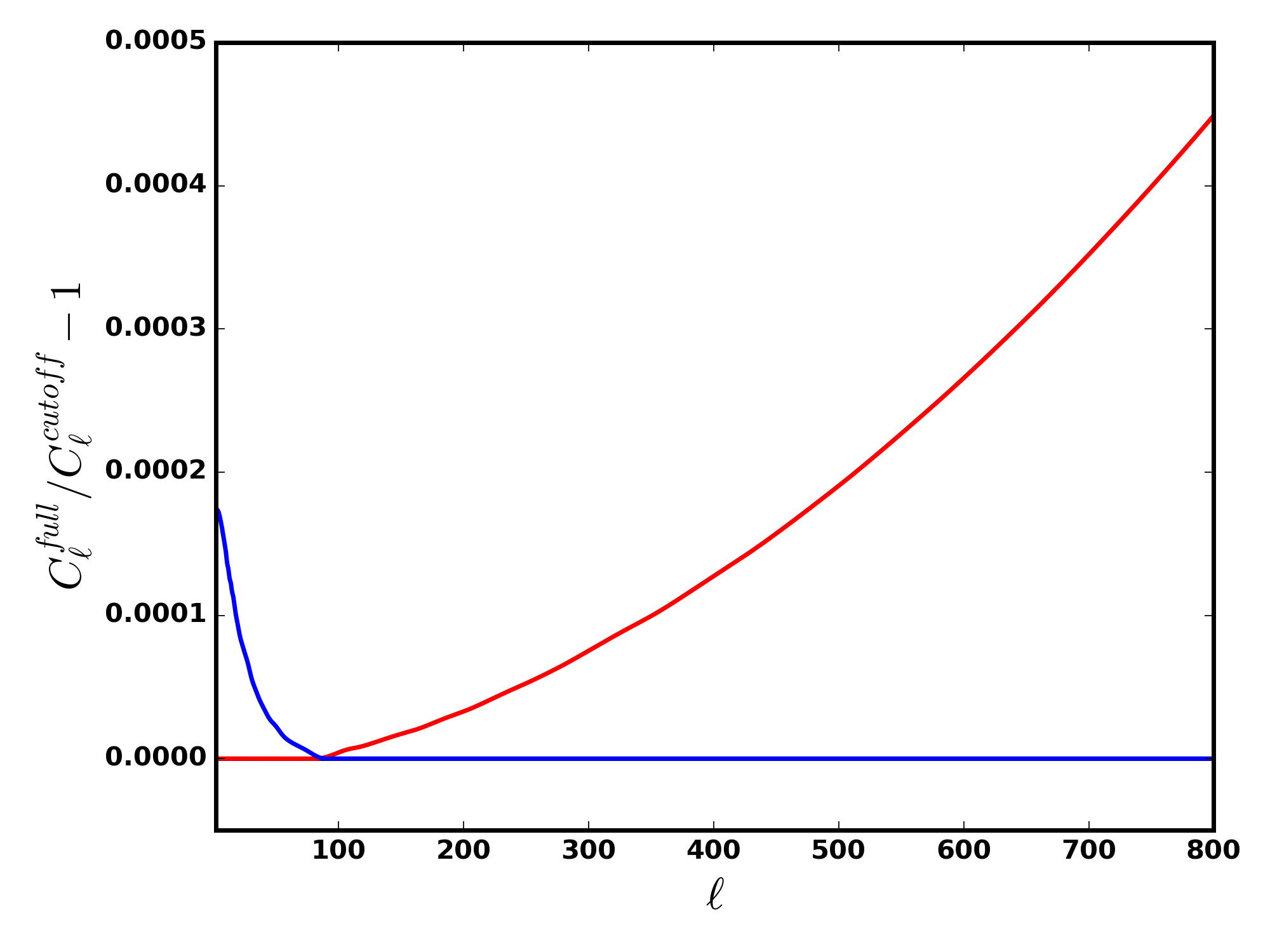}
\caption{Plots showing the fractional change on the CMB spectra when the $k$ (blue) and $z$ (red) cutoffs are turned off. The upper panel shows the ratio of the lensed temperature spectra, and the lower panel shows the ratio of the lensing potential spectra. In all cases the change to the $\Lambda$CDM spectrum is below 0.05\% for all scales of interest.
}
\label{fig_tt_kzcutoffs}
\end{figure}

Due to the computational expense of computing the halo model for all times and scales that are evolved in \texttt{class}, we implement $k$ and $z$ cutoffs on scales where the non-linearities are expected to be negligible. In particular, for $k$ such that the dimensionless power spectrum $\Delta^2(k,z)< A_{k_\text{cutoff}}$ we do not apply the corrections, and for $z>z_{\rm cutoff}$ we do not apply the corrections. For the work here, $z_{\rm cutoff}=3.0$ and $A_{k_\text{cutoff}}=0.001$. See figure \ref{fig_tt_kzcutoffs} for the numerical effect of removing these cutoffs on the temperature and lensing spectra: the effect is numerically negligible for the work here.

\begin{figure}
  \centering
 \includegraphics[width=\columnwidth]{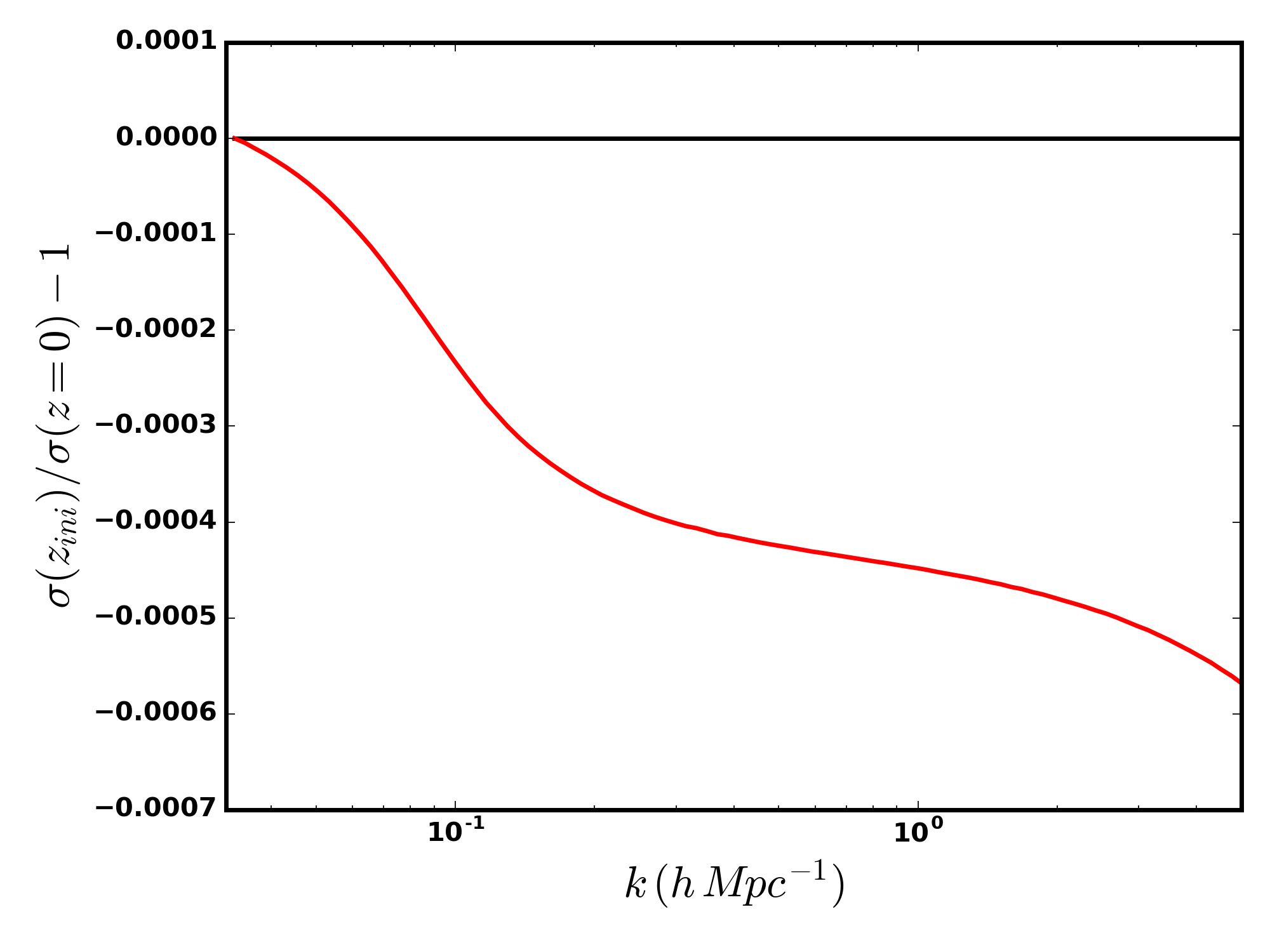}
\caption{Ratio of the smoothed power spectra, $\sigma_R(z_\text{ini})/\sigma_R(z=0)-1$, for $z_{\rm ini}=50$ calculated in \texttt{class}, as a function of smoothing scale $R$ for $\Lambda$CDM. The ratio is normalised to be unity at the lowest $R$ value. This shows the scale dependence of the growth due to the radiation contamination in the \texttt{class} code for $\Lambda$CDM. This scale dependence is below 0.05\% for all scales of interest here.
}
\label{fig_sigmaratio}
\end{figure}

As noted earlier, one of the subtleties of calculating the non-linear within \texttt{class} is that the radiation contamination at late times prevents the growth from being exactly scale invariant, even for $\Lambda$CDM. In figure \ref{fig_sigmaratio} we show the ratio between the smoothed matter power spectra at redshift zero and at our chosen value of $z_{\rm ini}=50$ for $\Lambda$CDM. This is normalised to be unity at the largest scale (lowest k-value). The scale dependence of the growth shown here is below $0.05\%$ and thus negligible for the purposes of the halo model calculations considered here.

\section{Use of the matter continuity equation for redshift space distortions and BAO reconstruction}
\label{sec_continuityappendix}
Both the use of redshift space distortions to calculate the growth of structure, and the use of reconstruction methods to improve recovery of the BAO peak, rely on the use of the continuity equation for the matter. This is something that is typically valid even in modified gravity theories, as it arises from the conservation of the matter energy-momentum tensor. However, once cold dark matter itself is changed, such as in the GDM model, then this relation no longer holds. Thus, CDM is assumed during such data analysis techniques.

In principle this could mean that our use of some BAO data for our constraints could be problematic,\footnote{Note however that one of the BAO datasets (6df) used in our constraints uses no reconstruction.} however we note that the expected effect should be small. Furthermore, the effect is expected to further improve our constraints, since it would increase the difference between GDM and $\Lambda$CDM, and thus the work here can be considered to be conservative. We now look briefly at the usual approach in order to see how the details might differ for GDM.

From number conservation, the real space density ($\delta_r$) and redshift space density ($\delta_s$) are related by 
\begin{equation}
\delta_s=\delta_r+\mu^2(-k^2 \theta)\text{,}
\end{equation}
where $\mu=\hat{r}\cdot\hat{k}$ is the angle between the wavevector and line of sight and $\vec{v}=-i\vec{k}\theta$, where $\vec{v}$ is the velocity perturbation of which $\theta$ is the scalar part. The standard continuity equation for pressureless CDM is given by
\begin{equation}
\label{eqn_cdmcont}
-k^2\theta=\dot{\delta}=\delta {\cal{H}} f \text{,}
\end{equation}
where the $f=d \ln \delta/d\ln a$ is the growth rate. Thus the difference between the real and redshift space densities is expressed as
\begin{equation}
\delta_s=\delta_r(1+\mu^2 {\cal{H}} f )\text{.}
\end{equation}
Instead, for GDM, the continuity equation is 
\begin{equation}
-k^2\theta=\frac{\dot{\delta}}{1+w}-k^2\zeta-\frac{1}{2}\dot{h}-\frac{3{\cal{H}}(w\delta-\Pi)}{1+w}
\end{equation}
Note that we are working in conformal time and $\dot{\delta}$ is the derivative with respect to conformal time.
Specifying the Newtonian gauge to remove $\zeta$, and dropping the time derivative term (as $\dot \Phi = -\dot h/6$ will be small), leaves us with the following replacement for equation \ref{eqn_cdmcont},
\begin{equation}
-k^2\theta=\frac{1}{1+w}\delta {\cal{H}} f-\frac{3{\cal{H}}}{1+w}\left(w\delta-\Pi \right) \text{.}
\end{equation}
Interestingly, since $\Pi$ is driven by $w$ and $c^2_s$, this relation might be useful as a way to break the $c^2_s-c^2_\text{vis}$ degeneracy if GDM model was fit directly to redshift-space correlation functions. We can consider several assumptions from here, in order to obtain an approximate value for the size of the correction. One could assume that the pressure $\Pi$ is small, in which case, if the growth rate is measured assuming CDM, then the quantity that has actually been measured ($f^{\rm \Lambda CDM}$) is 
\begin{equation}
f^{\rm \Lambda CDM } = \frac{f-3w}{1+w}\text{.}
\end{equation}
Alternatively, one could use $\Pi= c_s^2 \delta + 3 (c_s^2 - c_a^2) {\cal{H}} (1+w) \theta$ and assume that the non-adiabatic pressure vanishes (i.e. that $c_s^2=c_a^2$). In this case the quantity that has actually been measured is 
$\frac{f+3(c_a^2-w)}{1+w}$, which for constant GDM parameters is $\frac{f}{1+w}$. In either case, for values of $w$ consistent with our constraints from the Planck data, the difference between $f^{\rm \Lambda CDM }$ and $f$ is small (less than a percent level change), so we expect that our constraints are not sensitive to this effect.\footnote{However, note that relaxing the constancy of $w$ and allowing $w$ to vary freely with time, as done in \citep{KoppEtal2018}, increases the error bars on $w$ in the late universe by a factor 100 such that the modelling error in this case would not necessarily be small.}

\bsp	
\label{lastpage}
\end{document}